\documentclass[aps,prl,superscriptaddress,twocolumn,showpacs]{revtex4-1}

\usepackage[utf8]{inputenc}
\usepackage[english]{babel}
\usepackage{graphicx} 
\usepackage{amsmath}
\usepackage{amssymb}
\usepackage{xcolor}
\usepackage{color} 
\usepackage[abs]{overpic}
\usepackage{epstopdf}
\usepackage{soul}

\usepackage{ulem}

\definecolor{redred}{HTML}{D53E4F}

\definecolor{greengreen}{HTML}{2F9B31}

\newcommand{\ketbra}[2]{| #1 \rangle\langle #2 |}
\newcommand{\adag}[1]{a^{\dagger}_{#1}}
\newcommand{\bdag}[1]{\hat b^{\dagger}_{#1}}

\begin{document}

\title{Boundary time crystals}

\author{F. Iemini}
\affiliation{ICTP, Strada Costiera 11, 34151 Trieste, Italy}

\author{A. Russomanno}
\affiliation{NEST, Scuola Normale Superiore $\&$ Istituto Nanoscienze-CNR, I-56126, Pisa, Italy}
\affiliation{ICTP, Strada Costiera 11, 34151 Trieste, Italy}

\author{J. Keeling}
\affiliation{SUPA, School of Physics and Astronomy, University of St Andrews, St Andrews, KY16 9SS, United Kingdom}
 
\author{M. Schir\`o}
\affiliation{Institut de Physique Th\'eorique, Universit\'ee Paris Saclay, CNRS, CEA, F-91191 Gif-sur-Yvette, France}

\author{M. Dalmonte}
\affiliation{ICTP, Strada Costiera 11, 34151 Trieste, Italy}

\author{R. Fazio}
\affiliation{ICTP, Strada Costiera 11, 34151 Trieste, Italy}
\affiliation{NEST, Scuola Normale Superiore $\&$ Istituto Nanoscienze-CNR, I-56126, Pisa, Italy}

\date{\today}

\begin{abstract}
In this work we introduce {\it boundary time-crystals}. Here  {\it continuous} time-translation symmetry breaking occurs only  in a macroscopic fraction
of a many-body quantum system.  After introducing their definition and properties, we analyse in detail a solvable model where an accurate scaling 
analysis can be performed. 
The existence of the boundary time crystals is intimately connected to the emergence of a time-periodic steady state  in the thermodynamic limit of a many-body 
open quantum system. We also discuss connections to quantum synchronisation. 
\end{abstract}


\maketitle

{\it Introduction - } Spontaneous symmetry breaking is a cornerstone of physics and occurs at all energy scales, in cosmology and 
high-energy physics as well as in condensed matter. Thermal or quantum fluctuations can drive a system into a state that breaks, in 
the thermodynamic limit, some of the symmetries present in its (thermo)-dynamical 
potentials~\cite{Goldenfeld-book,Sachdev-book}.  
Can time-translation invariance be spontaneously broken? The possible existence of 
time crystals, first addressed by Wilczek in 
\cite{wilczek_2012}, prompted an intense 
discussion~\cite{li_2012,bruno_2013,nozieres_2013,volovik_2013}. 
A no-go theorem~\cite{watanabe_2015} ruled out the existence of time-crystals 
in thermal equilibrium in cases for which the energy is 
the only constant of motion.
 The situation may be different in the presence of additional
extended conserved quantities~\cite{footnote.Huse},
such as in superfluids~\cite{Sinatra_2016} where time-crystalline behavior was 
discussed in~\cite{Svistunov_2017}.
Ordering in time can also occur, however,  under non-equilibrium
conditions (e.g. by preparing the system 
in an excited state~\cite{syrwid_2017}). 

An important step forward in our understanding of spontaneous time-translational 
invariance  has been achieved 
in~\cite{Sacha_2015,else_2016,khemani_2016a} where Floquet time crystals, a.k.a. $\pi$-spin glasses, were introduced. The dynamics of 
these systems, subject to a periodic driving, is characterised by observables which oscillate at a multiple of the driving period. Hence they break the
{\it discrete} time-translation symmetry imposed by the external drive. Floquet time crystals were intensively explored from a theoretical point of view 
in~\cite{von-keyserlingk_2016_a,von-keyserlingk_2016_b,else_2017,yao_2017,ho_2017,huang_2017, russomanno_2017} and very recently experimentally 
observed~\cite{zhang_2017,choi_2017}.  A comprehensive review on time crystals can be found in~\cite{sacha_2017}.

\begin{figure}[t!]
\hspace*{-1cm}\includegraphics[width=0.42\textwidth,angle=-90]{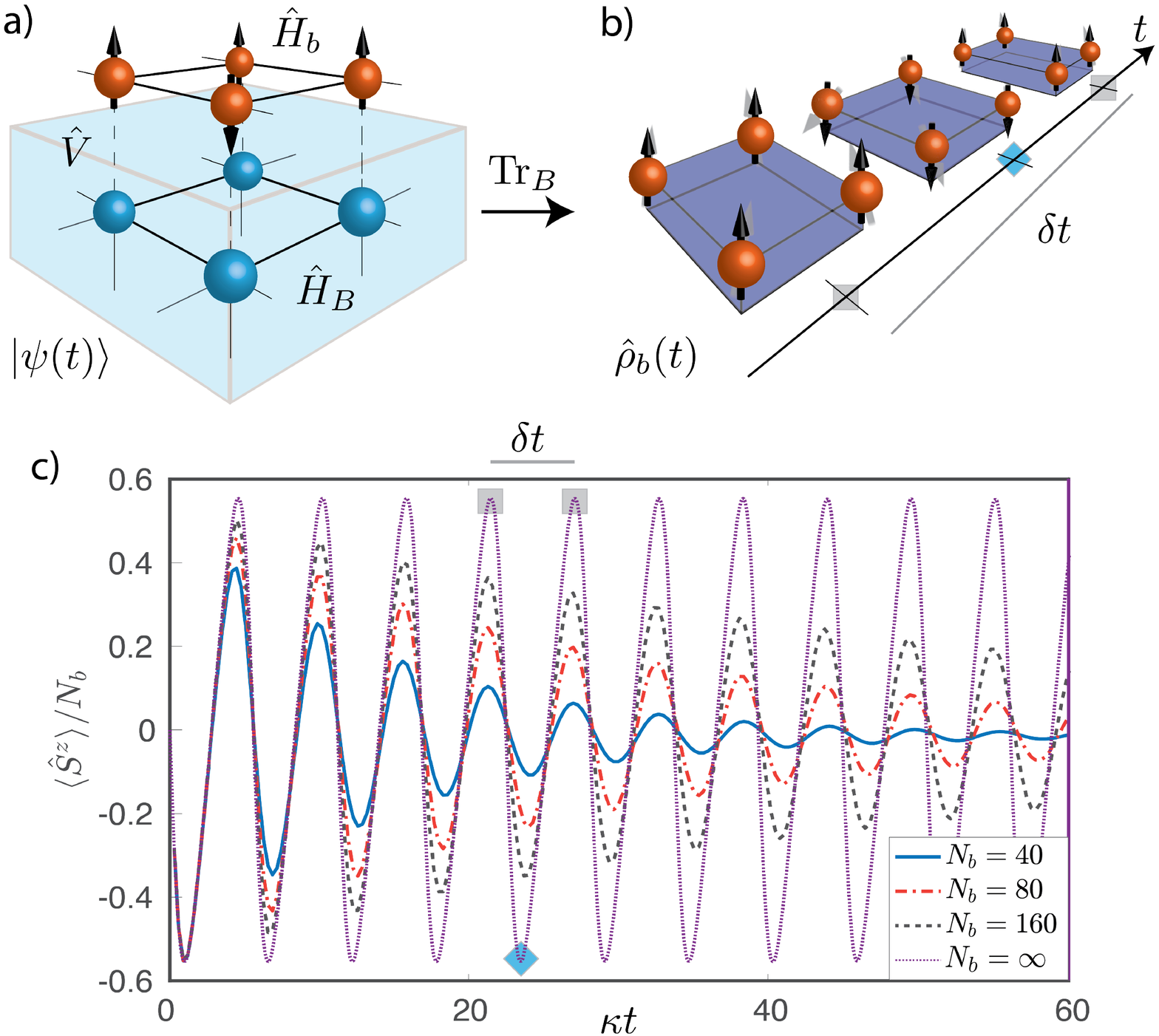}
\caption{ (Upper panel) A sketch of a boundary time crystal.  The system is composed by a bulk  (B) and a boundary (b) interacting trough an 
	interaction term $\hat V$. The Hamiltonian of the system is time-independent. After tracing 
	out the degrees of freedom of the bulk, the dynamics of the boundary is described by the reduced density matrix $\hat \rho_{\rm b}$. In the 
	time-crystal phase, the behaviour of collective variables will show persistent oscillations in the thermodynamic limit.
	(Lower panel) The boundary magnetisation of the  BTC model discussed in the paper is shown as a function of time, for 
	different boundary sizes. In the asymptotic condition, spontaneous symmetry breaking appears in persistent oscillations when $N_{\rm b} \rightarrow \infty$. }
\label{fig1}
\end{figure}

Here we predict a novel form of time-translation symmetry breaking: {\it continuous  Boundary Time-Crystals} (BTCs).  
In the  BTC phase, symmetry breaking appears in a (macroscopic) fraction of the system.  Moreover a BTC breaks the  continuous 
time-translation symmetry, i.e. the system self-organises in a time-periodic pattern with a period which only depends on its coupling constants. 
The idea borrowed from surface critical phenomena~\cite{binder-rev} offer a very intuitive way to visualise BTCs. Only the surface,
representing the portion of the system where time crystalline behaviour appear, is ordered. The rest of the system, the bulk
remains time-translationally invariant, see Fig.~\ref{fig1}. We will give a more precise meaning to this picture in the following of the paper where we will
show that BTCs are intimately connected to the existence of periodic motion in the steady state of open quantum many-body systems.

{\it Boundary time crystals - }  The emergence of a BTC can be understood using the sketch given in Fig.~\ref{fig1} (top panels).  A $d$-dimensional 
quantum many-body system is governed by a time-independent Hamiltonian $\hat H = \hat H_{\rm{B}}+\hat H_{\rm{b}}+\hat V$, with
bulk and boundary systems $\hat H_{\rm{B}}$ and $\hat H_{\rm{b}}$, respectively, and an interaction term  $\hat V$.
Denoting as $N_{b}$ $(N_{B})$ the degrees of freedom for the boundary (bulk) systems, we consider the case in
which a macroscopic fraction of the universe, the system ($N_b \rightarrow \infty$), breaks
spontaneously time-translational invariance. The thermodynamic limit is performed with $N_{b}$, $N_{B} \rightarrow \infty$, with the 
ratio $N_b/N_B \rightarrow 0$. In other words, it is a macroscopic system, but still small/infinitesimal compared to the global system. This scaling 
is the crucial feature in defining a boundary phenomenon.
The precise identification of the boundary layer (\textit{e.g.} the nature or any notion of spatial locality for its degrees of freedom) is thus irrelevant 
for our purposes.  The whole system evolves according 
to the Schr\"odinger equation $| \psi(t) \rangle = e^{-i \hat H t} | \psi(0) \rangle$, with  $| \psi(0) \rangle $ the initial state of the quantum 
system.  The boundary is fully characterised by the reduced density matrix $\hat \rho_{\rm{b}} = {\rm Tr}_{\rm{B}}\left( 
\ketbra{\psi(t)}{\psi(t)}\right)$ obtained by tracing out the bulk degrees of freedom. Its dynamics is governed by a completely 
positive, trace-preserving, map $\mathcal{\hat L}$ with
\begin{equation}
 	\frac{d}{d t}  \hat \rho_{\rm{b}} = \mathcal{\hat L}\left[\hat \rho_{\rm{b}}\right] \;.
\label{map}
\end{equation}
Time-translation symmetry breaking at the boundary appears as  a non-trivial time-dependence of  a (macroscopic) boundary order parameter  
$\hat O_{{\rm b}}$, occurring  only in the thermodynamic limit. For infinitely large times  its expectation oscillates, $\lim_{N_{\rm b} , N_{\rm B} \rightarrow \infty} {\rm Tr} [ \hat O_{{\rm b}} 
\hat \rho_{\rm{b}} ]  = f(t)$ where $f(t)$ is a time-periodic function. The definition of BTC  closely follows the one for the standard time 
crystals~\cite{watanabe_2015,von-keyserlingk_2016_a,von-keyserlingk_2016_b}. The only, crucial, difference is that here 
the order parameter is defined at the boundary.  

The resulting physical picture is exemplified in Fig.~\ref{fig1}, taking a magnetic system as an illustration.  In this example a macroscopic  
magnetisation builds up at the surface of a sample. The magnetisation shows persistent oscillations even though the dynamics of the 
whole system is governed by a time-independent Hamiltonian. In Fig.~\ref{fig1} the boundary and the bulk are represented with different 
symbols in order to stress that they may be described by different degrees of freedom. 
Notice that the terms bulk and boundary are used here to easily visualise the mechanism of spontaneous symmetry breaking and suggest an 
intriguing connection with boundary critical phenomena. What is really implied in the construction above is that ordering in time occurs only in a 
macroscopic fraction of the many-body system under consideration, rather than in the whole bulk.  

The boundary nature of time-translation symmetry breaking in BTC  has a number of important implications.  
First of all, the reduced density matrix $\hat\rho_{\rm{b}}$ in the steady state will be generically non-thermal, 
hence the no-go  theorem~\cite{watanabe_2015} does not apply: 
a Hamiltonian system can spontaneously break time-translation symmetry as a boundary phase. 
Furthermore, given the well known correspondence of the dissipative dynamics in Eq.(\ref{map}) and a 
unitary dynamics governed by a time-independent Hamiltonian on an enlarged system (see e.g.~\cite{nielsen}), the BTC appears tightly linked 
to the existence of a time-periodic steady state in an open quantum many-body system, appearing though {\it only} for $N_{\rm{b}} \rightarrow \infty$.
In order to discuss concrete examples we focus on boundary systems described by Markovian maps, and comment further below about more general dissipative maps.

The evolution of the boundary in the Markovian case  is described by a master equation where the Liouvillian operator $\mathcal{\hat L}\left[\cdot\right] $ has Lindblad 
form~\cite{nielsen},  $ \mathcal{\hat L}[\cdot] = \sum_{\alpha} \left\{\hat  \ell_{\alpha} \cdot \hat \ell_{\alpha}^\dagger - \frac{1}{2} 
\{ \hat \ell^\dagger_{\alpha}\hat \ell_{\alpha},\cdot \} \right\}$ with $\hat \ell_{\alpha}$ the Lindblad operators~\cite{footnote2}.
The emergence of a  time-crystal behaviour  in the  long-time dynamics of the system is hidden in the properties of the Liouvillian operator in the thermodynamic 
limit. In the BTC phase one should expect: i) a  vanishing gap in the real part of the Liouvillian eigenvalues,  making the non-equilibrium steady state subspace
degenerate in the thermodynamic limit with time-dependent coherences decaying over an infinite time-scale;   ii) a non-zero imaginary part for some Liouvillian 
eigenvalues in such subspace in order to induce non-trivial oscillations. The main question now is to 
find a many-body system that displays the above mentioned properties. Below we will present a  model of a BTC. 

{\it A BTC model  - }  We will show that a boundary time crystal appears in a model used to describe cooperative emission in 
cavities (see~\cite{hannukainen_2017,walls_1978,drummond_1978,puri_1979,walls_1980,schneider_2002}).  The boundary Hamiltonian  $\hat H_{\rm{b}} = \omega_0 \sum_j \hat \sigma^x_j$ 
consists in a collection of 1/2-spins whose dynamics is governed by collective spin operators $\hat S^\alpha = \frac{1}{2}\sum_j \hat \sigma^\alpha_j$. 
The operators $\hat \sigma_j^\alpha$ ($\alpha=x,y,z$) are the Pauli matrices acting on the j-th spins, and $\omega_0$ is the coherent splitting. 
The terms $\hat H_{\rm{B}}$ and $\hat V$ (see the sketch in Fig.~\ref{fig1}) have to be constructed in such a way to give a reduced dynamics at the 
boundary of the form 
\begin{equation}
 	\frac{d}{d t}  \hat \rho_{\rm{b}}  =   i \omega_0  [ \hat \rho _{\rm b},\hat S^x]
  	+ \frac{ \kappa}{S}\left(\hat S_- \hat \rho_{\rm b} \hat S_+ - \frac{1}{2}\{\hat S_+ \hat S_-,\hat \rho_{\rm b} \}\right) \;.
\label{master}
\end{equation}
In the previous equation, the collective raising/lowering spin operators are given by $\hat S_{\pm} = \hat S^x \pm \hat S^y$, $\kappa$ is the effective 
decay rate, and $S=N_{\rm b}/2$  is the total spin.  In the following the expectations of the observables are indicated as $\langle \cdot \rangle = {\rm Tr} [  
\cdot \hat \rho_{\rm{b}} ]$. 

The specific form of $\hat H_B$ generating the dynamics in Eq.~(\ref{master}) will play no role.  It  is 
possible to derive it~\cite{footnote2}. In the Supplementary 
Material we discuss in details how such a construction can be made~\cite{SI}. Moreover similar 
Liouvillian dynamics have been extensively considered in the context of atomic systems coupled to cavity modes.
Typically, the scenario in which a model such as Eq.~(\ref{master}) arises involves a system periodically driven at a finite frequency, with a time dependent Hamiltonian. Depending on the specific driving, such an explicit time dependence can be usually gauged away: one can define a Hamiltonian leading to Eq.~(\ref{master}) which is time-independent in some specific choice of frame.  As long as such a Hamiltonian exists, and is physical, our interpretation of the time-translation symmetry breaking as a boundary phenomenon of a closed quantum system is reasonable (see~\cite{SI} for a detailed discussion). Moreover -- as we are going to 
show -- the BTC shows a time-dependent pattern whose period solely depends on the coupling constants of the system and which is in general incommensurate with the driving period: the system breaks a continuous symmetry, rather than a discrete one. 
The BTC is in apparent contradiction with the expectation that the  density matrix of a system in 
contact with a single thermal reservoir attains a time-independent steady state~\cite{cresser_1992}. The solution to this apparent paradox lies in the diverging boundary size, $N_b\to \infty$, which
leads to a divergent decay time-scale for oscillations (see~\cite{SI}), as we better discuss below.

The steady state diagram of the model has two distinct phases~\cite{hannukainen_2017}. For $\omega_0/\kappa < 1$,  the total magnetisation 
is finite $\langle \hat S^z \rangle $.  In the opposite case,  $\omega_0/\kappa > 1$, all spins  align along the $x$-direction.  More details are reported in~\cite{SI}.

 \begin{figure}[h]
 \centering
 \begin{tabular}{cc}
 \hspace{-0.4cm} \includegraphics[width=0.5\textwidth]{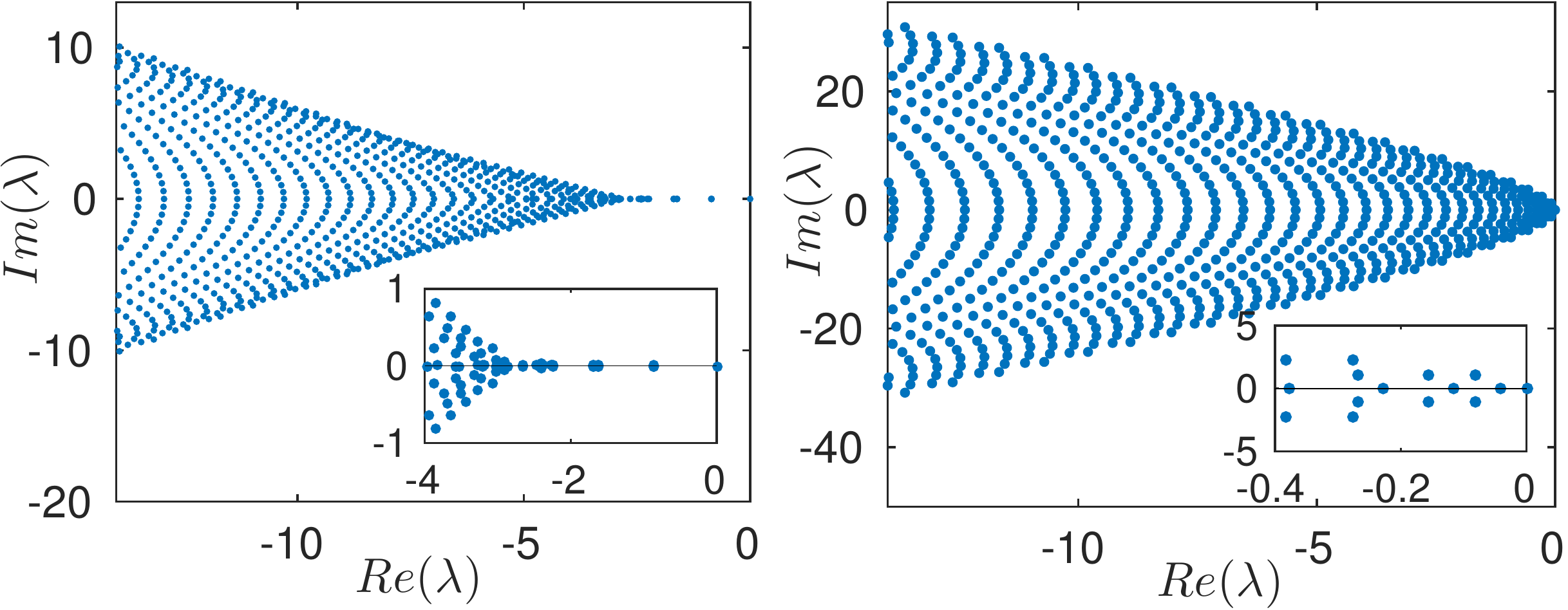}
 \end{tabular}
 \caption{The eigenvalues $\lambda$ of the Liouvillian  are shown in the weak dissipative case ($\omega_0/\kappa = 1.5$ -- right panel)
and in the strong dissipative one for ($\omega_0/\kappa = 0.5$ -- left panel), in a system with $N_{\rm b}=36$ spins. 
The insets show a zoom over the eigenvalues with largest real part. The eigenvalues are plotted in units of $\kappa$.}
\label{fig.structure.Liouv.spec}
 \end{figure}

The BTC appears for $\omega_0/\kappa > 1$.  Its emergence is embedded in the properties of the eigenvalues $\lambda$ of the Liouvillian
$\mathcal{\hat L}$. The structure of the Liouvillian spectrum is indeed different in the two phases. While for $\omega_0/\kappa<1$ the 
spectrum is gapped (Fig.~\ref{fig.structure.Liouv.spec} left panel), and the eigenvalues with greatest values for their real part (\textit{i.e., } the  eigenvalues closest to zero, recalling that $Re(\lambda_{j}) \leq 0$) have no imaginary values, for 
$\omega_0/\kappa>1$ the spectrum becomes gapless and the eigenvalues with greatest real part have a non zero imaginary part (see Fig.~\ref{fig.structure.Liouv.spec}).
The insets  zoom on the spectrum emphasising the different behaviour in the two limits.

\begin{figure}
 \centering
 \includegraphics[width=\columnwidth]{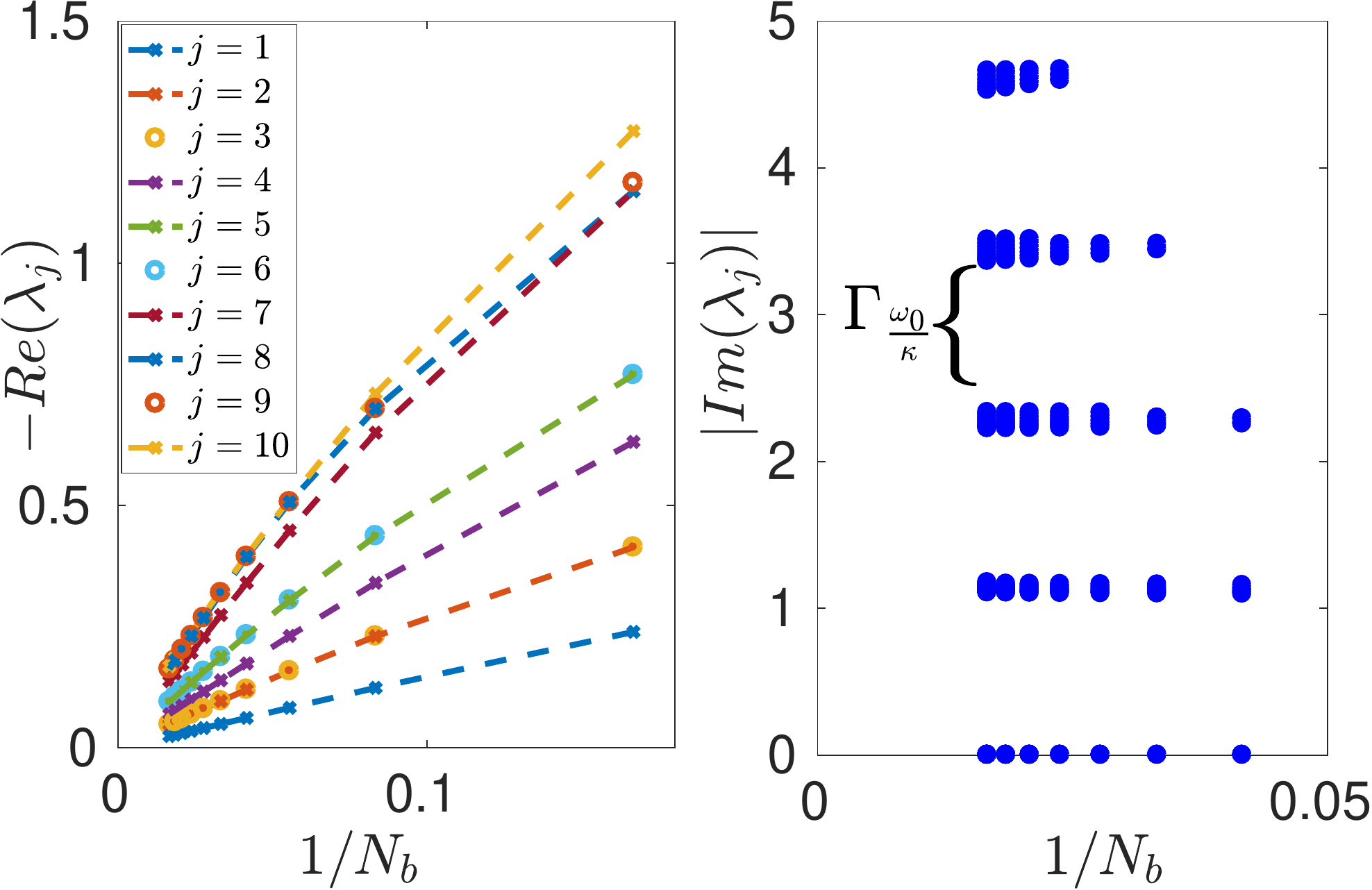}
 \caption{(Left) Finite size scaling for the real part of the Liouvillian eigenvalues in the BTC phase. The index $j$ labels the eigenvalues. The Liouvillian 
 	eigenvalues $\lambda_j$ are ordered as a function of their real part ($|Re(\lambda_j)|\leq |Re (\lambda_{j+1})|$, and $j=0$ has zero real part).  In the  $\omega_0/\kappa>1$ phase  they scale to 
	zero as a power-law of the inverse system size. (Right) The imaginary parts of the eigenvalues show a band structure, with a fundamental frequency 
	separation $\Gamma_{\omega_0/\kappa}$. For fixed excitation thresholds (we only select $\lambda_j$ such that $\nu =j^2/N_{\rm b} \leq \epsilon$) the width of the bands remains finite in the thermodynamic 
	limit (here we choose $\nu<0.025$). The widths of the bands tend to decrease as we constrain to lower excitation thresholds. The eigenvalues are plotted in units 
	of $\kappa$.}
\label{fig.fss.liouv.spec.part}
 \end{figure}

In order to obtain a quantitative picture of the development of the spontaneous symmetry breaking  we  perform a finite-size scaling analysis 
 of the real and imaginary parts of the eigenvalues $\lambda$. In Fig.~\ref{fig.fss.liouv.spec.part} (left panel) we analyse the real part of the Liouvillian 
spectrum. In the weak dissipative case, the one of interest to us, the system is gapless, with 
the real part of the eigenvalues closing with the system size as a power law (at different rates). In Fig.~\ref{fig.fss.liouv.spec.part} (right panel) we show the imaginary part of the Liouvillian spectrum. 
The imaginary eigenvalues of the low Liouvillian excitations are described by bands, separated by a fundamental frequency $\Gamma_{\omega_0/\kappa}$, 
which depends on the system parameters $\omega_0/\kappa$. These features of the real and imaginary parts are the key elements for the appearance of 
the BTC.

\begin{figure}[h]
 \centering
 \includegraphics[width=\columnwidth]{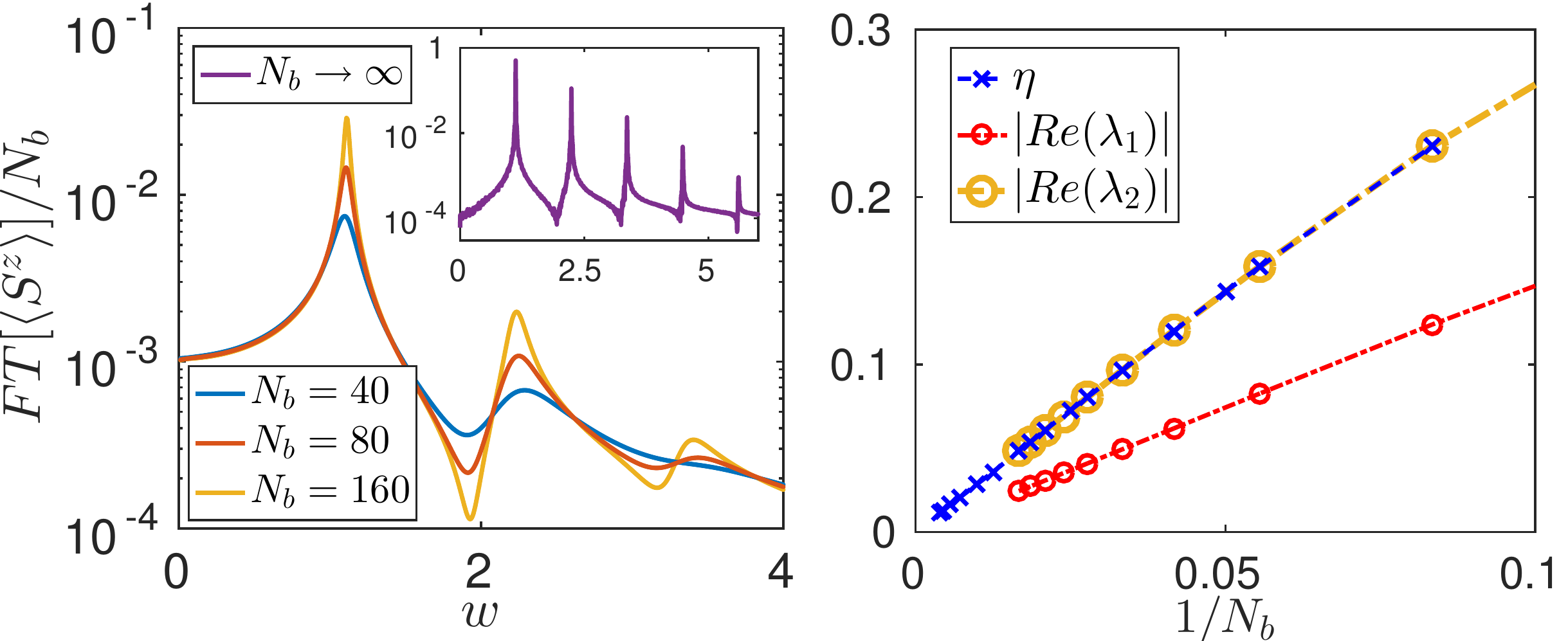}
 \caption{  In the right panel we plot the decay rate of the oscillations of the magnetisation $\eta$ for distinct system sizes. In the same plot we compare  
 	$\eta$ with the eigenvalues with greatest real part - thus smallest absolute values for the real part - of the Liouvillian.
 	In the left panel we plot the Fourier transform of the average magnetisation (see lower panel in Fig.~\ref{fig1}), highlighting the  oscillation 
	frequencies of the dynamics. The peaks are associated to the band separations in the imaginary part discussed in Fig.~\ref{fig.fss.liouv.spec.part}.
	The inset of the left panel is the solution in the thermodynamic limit where the oscillations persist indefinitely. 
  }
\label{fourier}
 \end{figure}

The magnetisation, for different numbers of lattice sites, is plotted in Fig.~\ref{fig1}
(lower panel). The system is initialised in the pure state with all spins  aligned along the $x$-direction. The oscillations decay for any finite size system, the associated time scale grows
 with the system size and 
diverges in the thermodynamic limit. This behaviour is independent of the initial conditions, as e.g., starting from thermal states or all spins aligned in different directions. Interestingly, the decay rate of the oscillations $\eta$ is related to the second excited eigenvalue of the Liouvillian, in our case,
the lowest excited eigenvalue with non zero imaginary value (eigenvalues are ordered according to the absolute value of their real part). A quantitative analysis of the spontaneous symmetry breaking is obtained by looking 
at the Fourier transform of $\langle S^z(t)\rangle$ (see Fig.~\ref{fourier}). By performing a  spectral analysis, we
see that the peaks appear at frequencies related to the separation between the bands shown in the right panel of Fig.~\ref{fig.fss.liouv.spec.part}. The peaks become sharper as
the system size is increased. Most importantly, the decay rate $\eta$ goes to zero (right panel) as a power law $L^{-\beta}$ with the $\beta$ exponent dependent on the system parameters $w_0/\kappa$. 
The finite-size scaling  shows that the persistent oscillations are associated to the spontaneous time-translation symmetry breaking because they
occur {\it only} in the thermodynamic limit. In the example we have discussed, the thermodynamic limit incidentally coincides with an effective classical dynamics (the 
effective Planck's constant going to zero): this is true for instance in  Fig.~\ref{fourier} and in 
Fig.~\ref{fig1} when $N_{\rm b} \rightarrow \infty$; details of the classical solution are 
discussed in~\cite{SI,note}.

From the experimental point of view, a driven version of this model can be realised using an adapted Raman
driving scheme~\cite{Dimer_2007} for cold atoms in an
optical cavity, connecting two low lying states via an excited atomic
state.  Collective dissipation can be produced by using a bad cavity
(large loss rate) combined with a single Raman drive: Purcell-enhanced
Raman scattering leads to optical pumping of the atoms, described by the
same collective dissipation considered here.  Similarly, the
Hamiltonian term $\hat{S}^x$ can be
realised by a pair of drive lasers coupling the ground states via excited states.

The boundary time-crystal we discussed in the model of Eq.(\ref{master}) is not an isolated point, but is robust to different perturbations. First of all, the time-crystalline 
phase appears in the whole region $\omega_0/\kappa > 1$. Moreover it is stable if additional perturbations are added to the unitary part of the 
evolution. With a boundary Hamiltonian of the  form 
 $ {\hat H}_{\rm b} = \omega_0  {\hat S}^x +  \omega_x  ({\hat S}^x)^2/S + \omega_z ({\hat S}^z)^2/S$, the time-crystal 
is still present for a wide range of the parameters $\omega_x,\omega_z\neq 0$. In fact the $\omega_x$ term 
improves the stability of the time 
crystal, which is also present for small values of $\omega_z$; 
for $\omega_z$ above some threshold, time-translation symmetry breaking still exists 
but only for some initial conditions (see~\cite{SI} for details). 
It is worth mentioning that robustness of a BTC phase refers to the persistence of a
periodic evolution in the thermodynamic limit, and not necessarily to the rigidity of its
period. The main difference with respect to Floquet systems is that there, since one is
breaking a discrete symmetry, rigidity is intimately related to the period of the driving;
instead, in our case, since the dynamics is $U(1)$ invariant, such timescale is not present,
and the period of oscillation is allowed to change within the symmetry-broken phase.
This is a direct analog of the fact that in spatial crystals, the spatial periodicity can be
changed by changing the particle-particle interaction.

It is also relevant to consider perturbations of the dissipative part of the evolution, more specifically we focus on terms non-local in time (this is equivalent to considering a non-Markovian equation of motion).  In order to have a physical 
bulk Hamiltonian, it must be bounded from below, and so it cannot have a truly flat density of states. This 
implies a finite memory timescale for the bath, but there is the possibility that this timescale can be neglected, 
being far smaller than all the other timescales in the system dynamics.
This fact occurs if the lower bound on the bulk spectrum is at energies much lower than the frequencies of the 
system dynamics: in this case an approximate Markovian description holds and the use of a Markovian master equation is perfectly justified.

{\it Other candidate systems for BTCs - } An interesting model that should show the same phenomenology has been studied 
in~\cite{hartmann_2016}. Furthermore, many-body limit cycles have been already seen in model systems of optomechanical arrays~\cite{ludwig_2013},  
coupled cavity arrays~\cite{cav_2013-2016_a,cav_2013-2016_b}, interacting Rydberg  atoms~\cite{lee_2011} and interacting 
spin-systems~\cite{chan_2015,chan_2015}.  Also in these cases the underlying (bulk+boundary) Hamiltonian can be constructed, see Ref.~\cite{SI}. 
In light of the analysis performed in the present work, these limit-cycles now might be classified as BTCs. It should be 
however kept in mind that a mean-field approximation, 
employed in these works, may be unable to support the very existence of limit cycles: 
it is not clear to which extend this phase would survive when fluctuations are included.

Other promising systems that it might be interesting to consider to seek for different forms of BTCs are dissipative topological systems.
In this case the steady state may develop a degeneracy in the thermodynamic limit due to the presence of edge states~\cite{diehl_2011,iemini_2016}. 
The existence of a BTC phase should emerge from the competition of the unitary and dissipative parts of the dynamics. Furthermore the 
robustness should be inherently linked to topological protection. 

Finally a BTC, corresponding to a space-time ordering, represents in essence a synchronised dynamics in a many-body open quantum system. This hints to 
a very interesting and deep connection between time crystals and quantum synchronisation. Lately there has been an intense effort to characterise 
synchronisation in the quantum realm (see e.g., the review \cite{galve_2016}).  BTCs may offer a different perspective on this problem. 

{\it Conclusions - } In this work we introduced  boundary time crystals. In the same spirit as in the original definition given in~\cite{watanabe_2015},
in the BTC phase the time-dependent order parameter appears only in a portion of the sample (at the boundary for simplicity). The phenomenon 
is analogous to surface critical phenomena. On looking at the reduced dynamics at the boundary, one observes that BTCs are intimately linked to 
the emergence of a periodic dynamics in some macroscopic observable of an open quantum many-body system. A crucial aspect of the 
whole picture is that the periodic motion should appear {\it only} in 
the thermodynamic limit.  We proposed an example of a BTC phase 
in a solvable model where its existence can be confirmed without resorting to any approximation. We finally discussed that BTCs can also emerge from 
different mechanisms in topological systems.

While completing this manuscript, a few works appeared 
\cite{Gong_2017,Wang_2017} analysing \textit{discrete} time crystal 
phenomena in periodically driven dissipative systems.

{\it Acknowledgements -}
We acknowledge enlightening  discussions with S. Denisov, J. Jin, L. Mazza, T. Prosen, and P. Zoller.  
This work was supported in part by ``Progetti Interni - Scuola Normale Superiore'' (A.R.), EU- 691 QUIC (R.F. and A.R.), CRF Singapore Ministry of 
Education (CPR-QSYNC 692) (R.F.), 
EPSRC program TOPNES (EP/I031014/1) (J.K.),

\clearpage
\newpage

\clearpage
\setcounter{equation}{0}%
\setcounter{figure}{0}%
\setcounter{table}{0}%
\renewcommand{\thetable}{S\arabic{table}}
\renewcommand{\theequation}{S\arabic{equation}}
\renewcommand{\thefigure}{S\arabic{figure}}

\newcommand{\ud}{\mathrm{d}}
\newcommand{\Tr}{\operatorname{Tr}}
\newcommand{\mean}[1]{\left\langle #1\right\rangle}
\newcommand{\atan}{\operatorname{atan}}

\onecolumngrid

\begin{center}
  {\Large Supplementary Information \\for\\\textbf{ Boundary time crystals}}
  
 \vspace*{0.5cm}
 F. Iemini, A. Russomanno, J. Keeling, M. Schir\`o, M. Dalmonte and R. Fazio
 \vspace{0.25cm}

 \end{center}

In this Supplementary Information we give more details on the model considered in Eq.~(2) of the main text, 
the structure of its Liouvillian spectrum, and provide a full discussion 
of the classical effective model for $N_b\to\infty$. We also analyse
the stability of the boundary time crystal when a perturbation of the type $(\hat{S}^z)^2$ is added to the Hamiltonian. For reader's convenience, we rewrite 
here the Lindblad equation
\begin{equation} \label{eq:master.equation.SI}
  \frac{\ud}{\ud t}\hat{\rho} = i[\hat{\rho},\hat H_{\rm b}]+\frac{\kappa}{S}\left(\hat{S}_-\hat{\rho}\hat{S}_+-\frac{1}{2}\left\{\hat{S}_+\hat{S}_-,\hat{\rho}\right\}\right)\,,
\end{equation}
where
\begin{equation}
	\hat H_{\rm b} =\omega_0\hat{S}^x+\frac{\omega_x}{S} (\hat{S}^x)^2
+ \frac{\omega_z}{S} (\hat{S}^z)^2
\end{equation}
and $N_b=2S$. All the results in the main text are for $\omega_x=\omega_z=0$. In the further sections we 
will show how they are extended to the more general case.
%

\section{From Hamiltonian dynamics to the Master equation}

In this section we add more details on the connection of Eq.~(2) of the main text
with an Hamiltonian dynamics 
and discuss how to apply a ``chain-mapping'' \cite{Plenio2010,Plenio2016} unitary transformation on the bath 
such that the bath have only local interactions and 
the system sits on its boundary. In this framework, we will also construct more 
general forms of boundary-bulk systems with the corresponding Hamiltonians.

We first compare 
two model Hamiltonians which lead to Eq.~(2) in appropriate limits.  A time-dependent
Hamiltonian leading to driven oscillations, typically appearing in problems of atoms collectively coupled to a 
cavity~\cite{walls_1978, walls_1980,drummond_1978,schneider_2002,hannukainen_2017_SI}, might take the form
\begin{equation} \label{h_bab:eqn}
  \hat H_b=\omega_0 \hat S^z + \hat S^x \cos(\omega_0 t),\quad \hat V=\hat S^x (\hat B+ \hat B^\dagger),
\end{equation}
where $\hat B$ is a combination of bath operators.  
For this Hamiltonian, the existence of a large drive frequency $\omega_0$ is necessary in order justify the 
approximation of neglecting fast oscillating terms 
in the system-bath Hamiltonian. 
However,
Eq.~(2) can also arise from a model in which counter-rotating terms are 
absent by definition (see for instance~\cite{walls_1980}),
\begin{equation} \label{h_bab1:eqn}
\hat H_b=\omega_0 \hat S^z + \left(\hat S^+ e^{-i \omega_0 t} + \text{H.c}\right), \quad
\hat V= \hat S^+ \hat B + \text{H.c.}\,.
\end{equation}
While such a model is not typically 
encountered, it remains physical (Hermitian, and bounded from below), and there exists
a frame in which the time dependence vanishes.  Even when considering 
this time-dependent Hamiltonian, we may note that the boundary time crystal corresponds 
to inducing a time dependence that is incommensurate with the driving term, hence 
the breaking of a continuous, rather than discrete symmetry.

We remark that the scenario we considered concerns an
optical frequency drive, for which a rotating wave
approximation (RWA) is valid. In this approximation one can perform a unitary transformation to a
frame in which the Hamiltonian is time independent. At lower frequencies the RWA cannot be made and there
is no frame in which the Hamiltonian is time independent. In these cases one can only define Floquet eigenvalues up
to an integer multiple of the drive frequency~\cite{sambe}, and the Floquet spectrum is
periodic in energy. In contrast, when there exists a frame for which the Hamiltonian is
time independent, it is possible to define eigenvalues in the ``standard'' way. The
difference between these two cases can directly be traced to the presence or absence of
counter-rotating terms in the effective Hamiltonian.

The physicality condition required can now be stated in a clearer way, in order to encompass
cases where the RWA is possible and cases where Floquet theory must be applied: the 
bath spectrum should be bounded from below. We state this condition in order
to exclude unphysical models where there is a continuous flow of energy into the bath
only because of its unbounded energy spectrum. A bath spectrum bounded
from below leads to non-Markovian effects at very-short time scales that do not affect the
existence of the time crystal.

\begin{figure}
\centering
 \includegraphics[width=0.8 \textwidth]{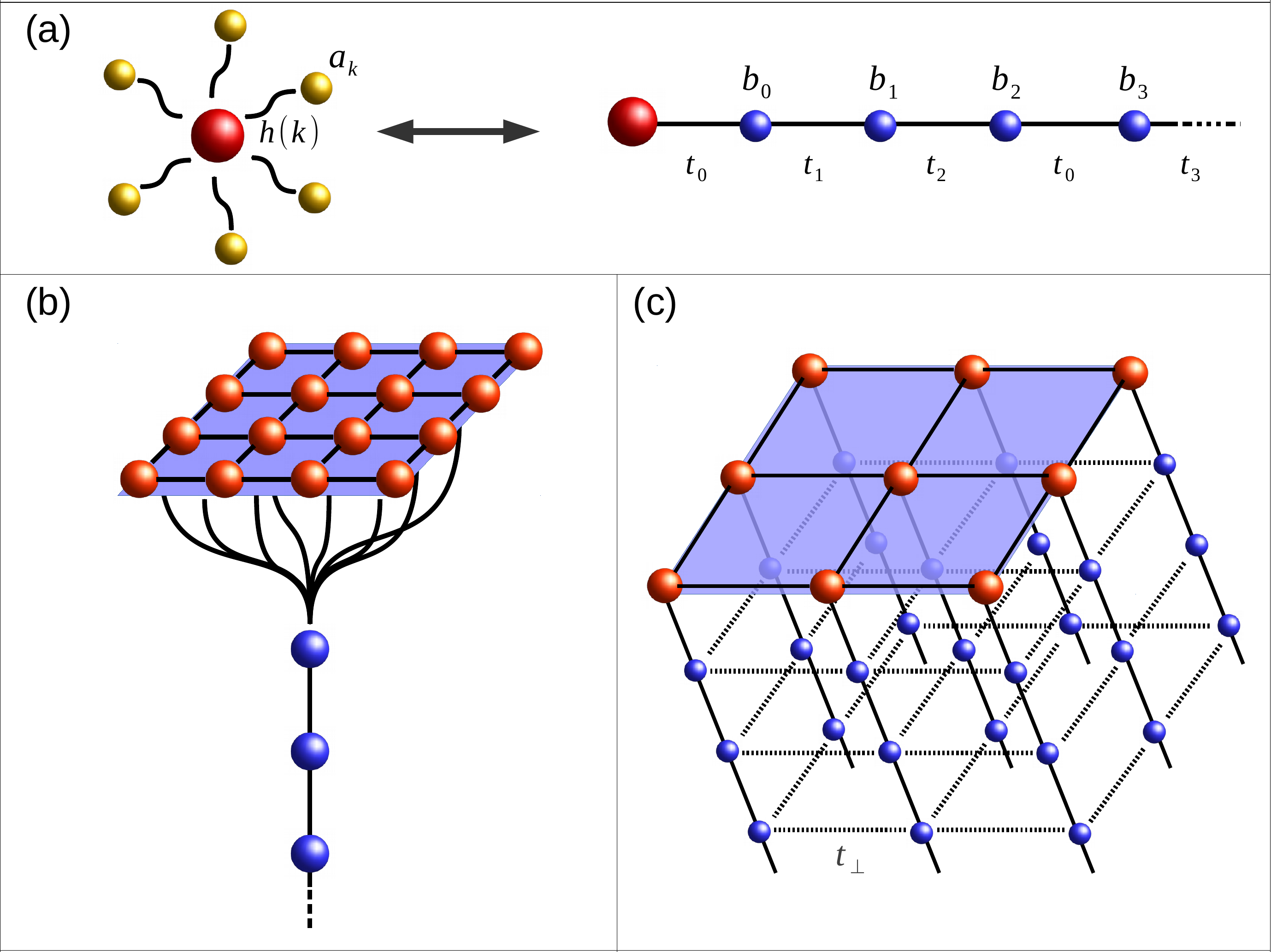}
 \caption{ \textbf{(a)} A sketch on the chain mapping transformation, in which a general 
 Hamiltonian system interacting 
 with an environment composed of independent harmonic oscillators, as represented by the ``star-shaped'' picture,
  is mapped to a one-dimensional chain with only local and nearest-neighbor interactions.
   \textbf{(b-c)} Two distinct boundary systems from the perspective of the chain mapping transformation: in 
   \textbf{(b)} the boundary system interacts collectively with the environment, while in \textbf{(c)} we consider
  the case in which all of the boundary elements interact with its own independent environment, 
  thus leading to a lattice structure with $t_\perp = 0$. Small interactions 
 $t_\perp \sim 0$ between nearby environmental modes shall not significantly influence 
 the boundary system dynamics, and we may even consider a full $(2+1)$-dimensional lattice structure in 
  these cases.
           } 
 \label{fig.chain.mapping}
 \end{figure}

\textit{Chain mapping:} In a more general form, one can define a boundary system and its corresponding Hamiltonian by using the chain 
mapping transformation, precisely, 
 an analytic transformation in which a general system interacting with an environment composed 
 of harmonic oscillators is mapped to a one-dimensional chain with only local and 
 nearest-neighbor interactions \cite{Plenio2010,Plenio2016}. The system after the 
 transformation is described as a
 boundary of the chain. The details of such transformation are given as follows.

Consider a general system linearly interacting with an environment 
 composed by independent harmonic oscillators. The total Hamiltonian can be described by,
 \begin{equation}
  \hat H = \hat H_S + \hat H_E + \hat H_{SE}
 \end{equation}
 where $\hat H_S$ describes the system Hamiltonian, $\hat H_E$ the environment, 
 and $\hat H_{SE}$ the system-environment interaction. Specifically,
 \begin{eqnarray}
  \hat H_E &=& \int_{0}^{k_{\max}} g(k) \,\adag{k}\hat a_k \,dk\\
  \hat H_{SE} &=& \hat A_S \int_{0}^{k_{\max}} h(k)  \left( \adag{k} + \hat a_k \right)dk
 \end{eqnarray}
 where $\adag{k} (\hat a_k)$ are bosonic creation (annihilation) operators for the environmental 
 modes, $g(k)$ its dispersion relation, $k_{\max}$ denotes
  the cut-off for the spectral density, 
 and the interaction is described by a general system operator $\hat A_S$ coupled
  to environment displacement operators with strength $h(k)$.

 Employing a bosonic basis transformation, one can map
 such ``star-shaped'' Hamiltonian to a
 one-dimensional chain with only local and nearest-neighbor
  interactions (see Fig.\eqref{fig.chain.mapping}-(a)). The system is mapped to a boundary of the chain.
  This transformation is accomplished by representing the bath with the new
   following set of bosonic operators:
  \begin{equation}
   \bdag{n} = \int_{0}^{k_{\max}} U_n(k) \,\adag{k} \,dk
  \end{equation}
  where $U_n$ are the coefficients of a unitary transformation based on orthogonal polynomials \cite{Plenio2010}.
 The Hamiltonian in this new basis is described by $\widetilde{H} = \hat H_S +
  \widetilde{H}_E + \widetilde{H}_{SE},$ with
  \begin{eqnarray}
   \widetilde{H}_{SE} &=& t_0 \hat A_S \left( \bdag{0} + \hat b_0 \right), \\
   \widetilde{H}_{E} &=& \sum_{n=0}^\infty w_n \bdag{n} \hat b_n + 
   t_n \left( \bdag{n} \hat b_{n+1} + h.c. \right),
  \end{eqnarray}
  where the effective hopping parameters $t_n$ are determined by the bath dispersion relation.

 The physical picture of a boundary system as defined in the main text then follows very naturally
  under such a chain-mapping transformation: in the limit 
  $N_B\to\infty$, $N_b\to\infty$, $N_b/N_B\rightarrow 0$ ($N_b$ and $N_B$ are the 
  system and 
   environment degrees of freedom, respectively) the bath behaves as 
   an infinite chain with the system always sitting on the boundary of this chain.

 In Fig.\eqref{fig.chain.mapping}-(b-c) we schematically illustrate two distinct 
 boundary systems under the perspective of the chain-mapping transformation:
 (i) in the first case, as shown in Fig.\eqref{fig.chain.mapping}-(b), 
 we consider all spins of the boundary system collectively interacting
  with a single bosonic mode of the transformed basis (the model studied in this article
   corresponds to this class, with a collective coupling 
   $\hat A_S \equiv \hat S_x$ -- see Eq.~\eqref{h_bab:eqn}); (ii) in a different setting, we can consider 
   the case where each spin of the boundary system interacts independently with its own environment. In the latter case, by
  applying the chain-mapping transformation to each spin, we get the lattice structure of Fig.\eqref{fig.chain.mapping}-(c).
 In both cases, under appropriate limits, such as weak coupling ``$t_0$'' between 
 system and environment, we can 
 neglect memory effects on the system dynamics, in the limit of an infinite chain. The dynamics
  of the system under these limits can be approximated by a Markovian dynamics, as discussed in detail
   at the beginning of this section for the specific model studied in this article.

 It is worth mentioning that the model considered in this article
 is particularly well suited for the study of BTC (boundary time crystal)
  due to the possibility of performing the scaling analysis (see Fig.2-4 in the main text).
  Time crystals are collective phenomena occurring
only in the thermodynamic limit and it is extremely important to perform a scaling
analysis in order to understand if this is the case.
 However, there are no arguments precluding the possibility of BTC's 
 in distinct systems, such as the lattice structure depicted 
 in Fig.\eqref{fig.chain.mapping}-(c).
  In fact, these models support limit cycles \cite{lee_2011_SM,chan_2015_SM} at the 
  mean-field level. In these cases however the problem is currently intractable and one is not able to
  derive an exact solution.
  More specifically, the models in \cite{lee_2011_SM,chan_2015_SM} are realised by having an XYZ-Heisenberg or quantum-Ising Hamiltonian at 
  the boundaries  (red sites in the figure) and local baths ($t_{\perp} = 0$). The 
  results are not 
  expected to change qualitatively at small $t_{\perp}$.

\section{Phase diagram and Liouvillian spectrum} 
In all this section we assume $\omega_x=\omega_z=0$.
The phase diagram of the model is relatively simple, with just two distinct phases 
according to the ratio $\omega_0/\kappa$: 
(i) if the dissipative part is the leading term ($\omega_0/\kappa < 1$ -- strong dissipative phase) 
all spins are aligned along the $z$-direction, and the total
 magnetization is finite $\langle \hat S^z \rangle <0$;
 (ii) on the other hand, if the driving Hamiltonian is the leading term ($\omega_0/\kappa > 1$ -- weak dissipative phase)
{the expectation values of all spins tend to be aligned along the $x$-direction  and now $\langle \hat S^z \rangle  = 0$.  
  In Fig.~\ref{fig.phase.diagram} we plot the expectation values of the collective spin
  operators (left panel) and their variances (right panel), highlighting the two distinct phases of the model.

 \begin{figure}
\centering
 \includegraphics[width=0.45\textwidth]{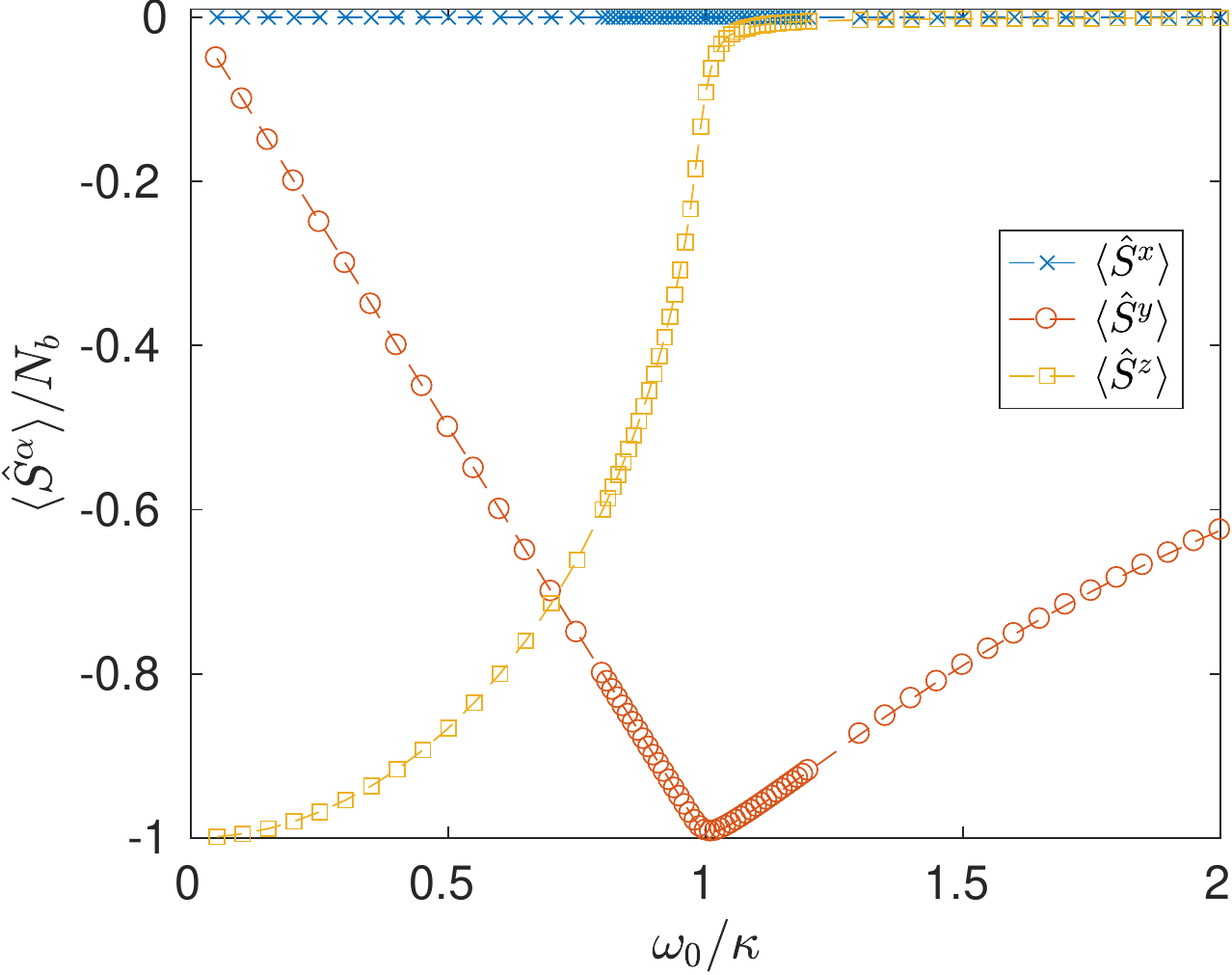}
 \includegraphics[width=0.45\textwidth]{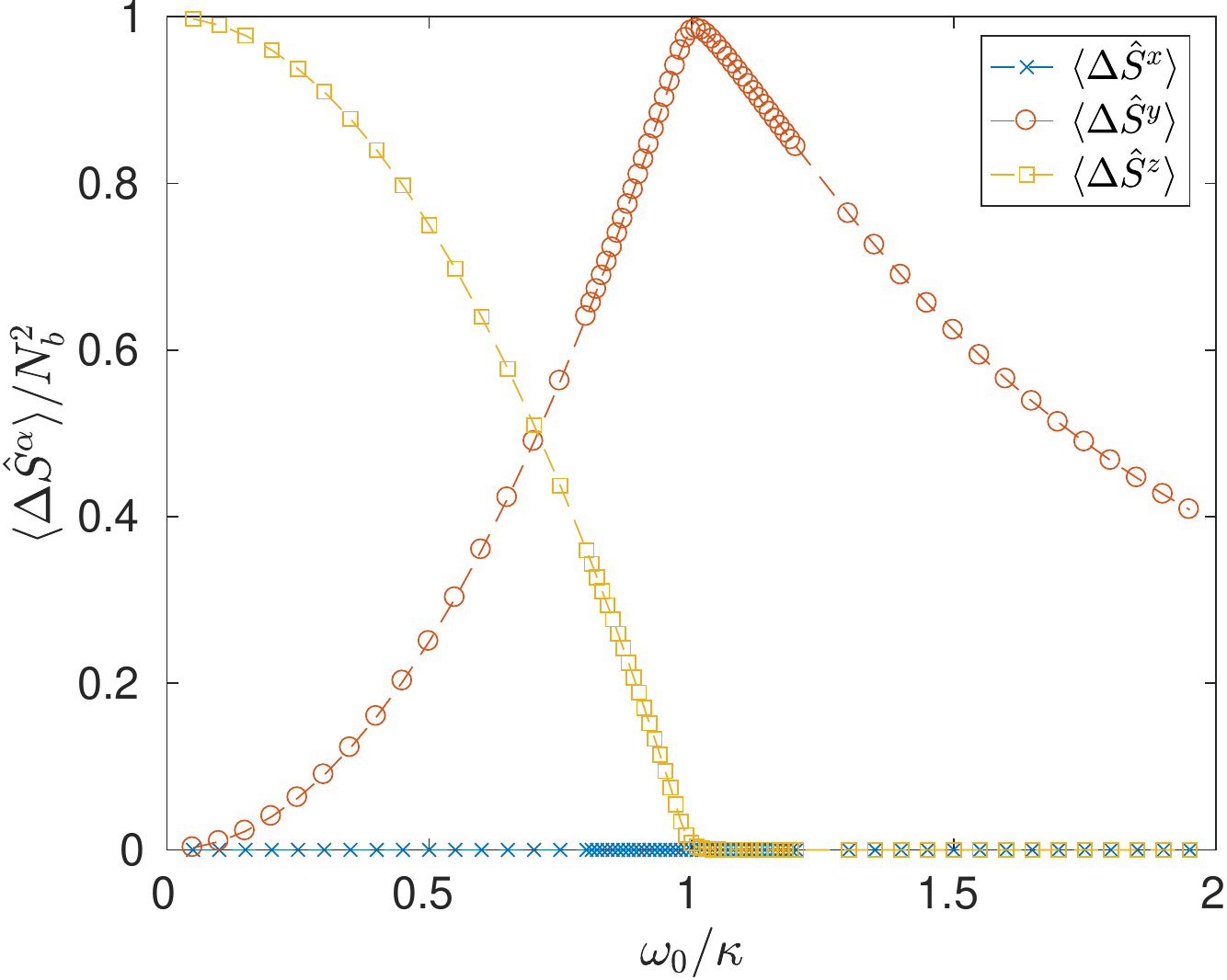}
 \caption{ Expectation value for distinct observables for the non equilibrium steady state (NESS) of the Liouvillian in Eq. (2) - this state is unique 
 for finite systems - in a system with $N_b=600$ spins. On the \textit{left panel} 
 we plot the expectation value for 
 the collective spins, as a function of the ratio $\omega_0/\kappa$,
 and on the \textit{right panel}  the expectation value of their variances. In both panels we have fixed $\omega_z=0$.}
\label{fig.phase.diagram}
 \end{figure}

As we show in the main
text, the properties of the Liouvillian spectrum are crucial for the time-crystal behaviour. In particular, in the weak dissipative phase, the gap in the real part of the Liouvillian spectrum vanishes in the thermodynamic limit, giving rise to the persistent time-translation symmetry breaking oscillations.
In Fig.~\ref{fig.fss.liouv.spec.realpart}, we show the finite size scaling for the real part of the Liouvillian eigenvalues in the two phases
  of the model. The Liouvillian eigenvalues $\lambda_j$ are ordered in terms of their real part: $|Re(\lambda_j)|\leq|Re(\lambda_{j+1})|$.
  In the upper panel the system is in the phase $\omega_0/\kappa<1$,
   and we clearly see a finite Liouvillian gap in the thermodynamic limit: no time-translation symmetry breaking occurs.
   On the opposite, in the lower  panel the system is in the phase $\omega_0/\kappa>1$, and 
   we see that the real part of the eigenvalues vanishes polynomially in $N_b$.
 \begin{figure}
 \centering
 \includegraphics[width=0.45\textwidth]{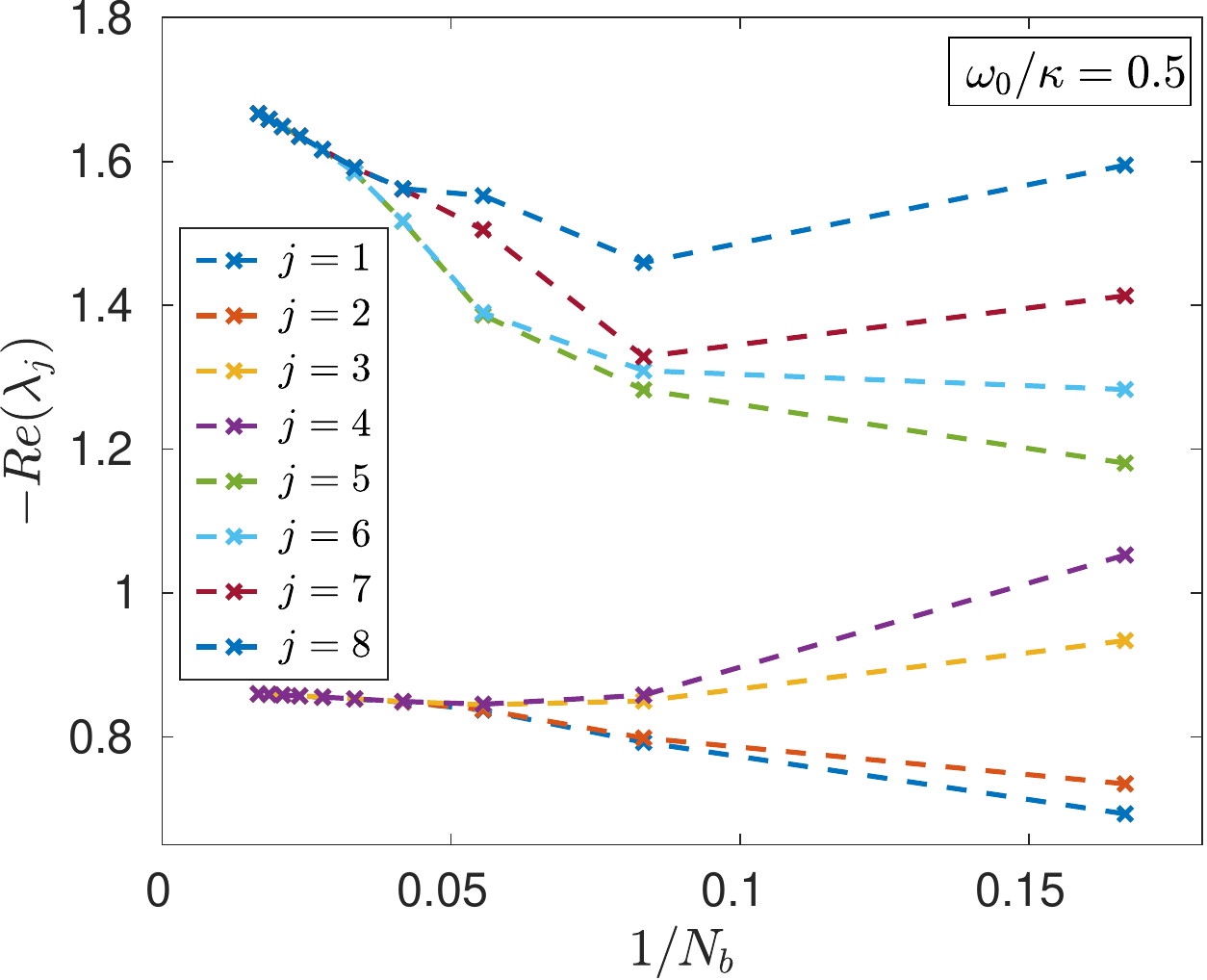}
 \includegraphics[width=0.45\textwidth]{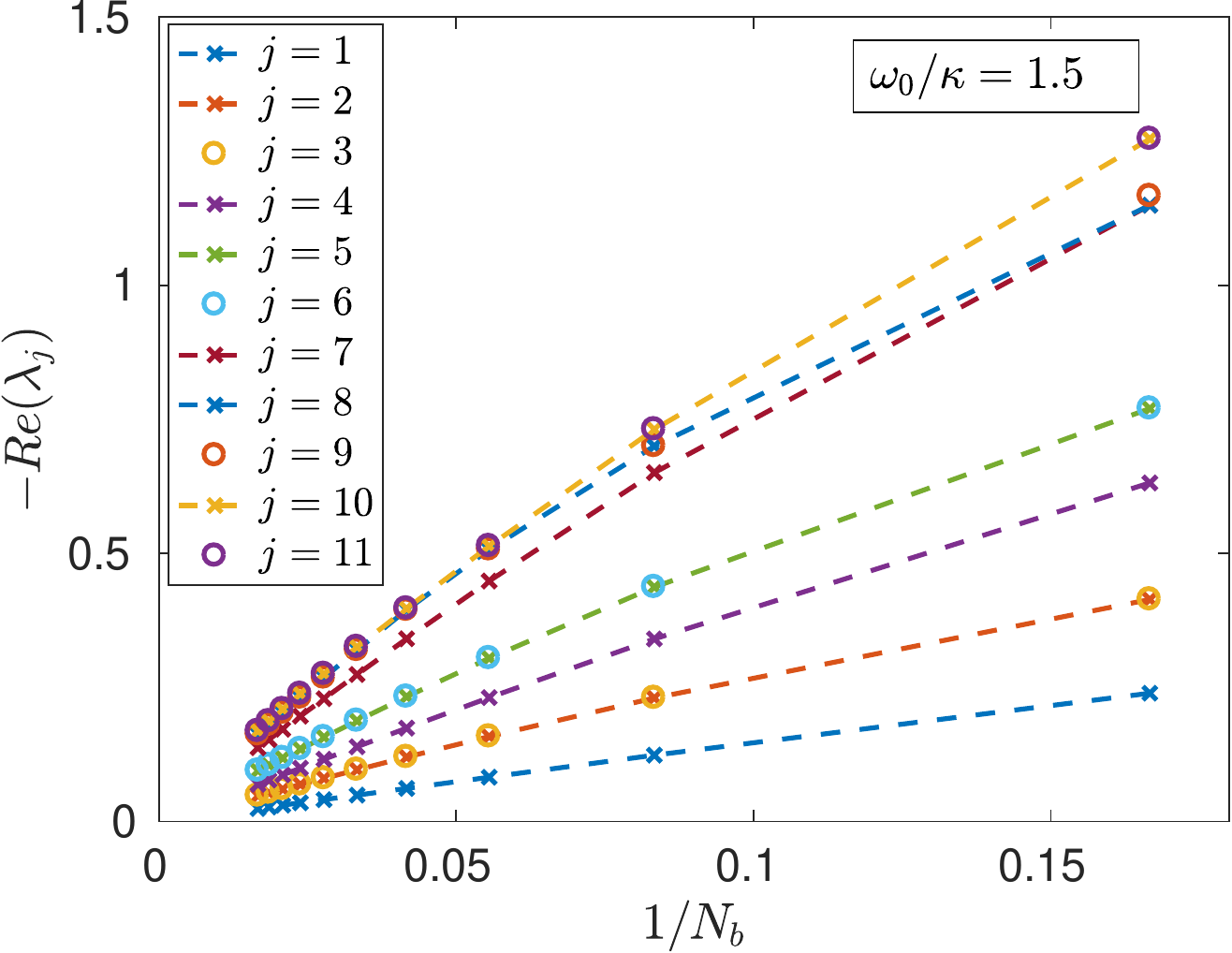}
 \caption{Finite size scaling for the real part of the Liouvillian eigenvalues in the two distinct phases
  of the model. (Left panel) For $\omega_0/\kappa=0.5$ the gap in the Liouvillian spectrum persist in the thermodynamic limit $N_b\to\infty$ while
for $\omega_0/\kappa=1.5$ (right panel) the eigenvalues vanish algebraically with $N_b$.
   }
\label{fig.fss.liouv.spec.realpart}
 \end{figure}

In order to underline the connection of the spectral properties with the time-translation symmetry breaking oscillations, in Fig.~\ref{fig.fss.liouv.spec.imagpart} we plot the absolute value of the lowest excitation Liouvillian
 eigenvalue with nonvanishing imaginary part, $\lambda^{Im}_<$.
  Remarkably, this value provides a good approximation to the fundamental frequency of the band 
  structure: $\lambda^{Im}_{<}/\Gamma_{\omega_0/\kappa} \sim 0.971 \,(0.995)$ for $\omega_0/\kappa = 1.5 (2)$ (see right panel of Fig.~3 of the main text).
   The fundamental frequency is here defined as 
the difference between the average position of two nearest bands. It also coincides with the frequency of the main peak in the Fourier transform of the oscillating time-translation symmetry breaking magnetization (see Fig.~4 of the main text).
%
  \begin{figure}
 \centering
  \includegraphics[width=0.45\textwidth]{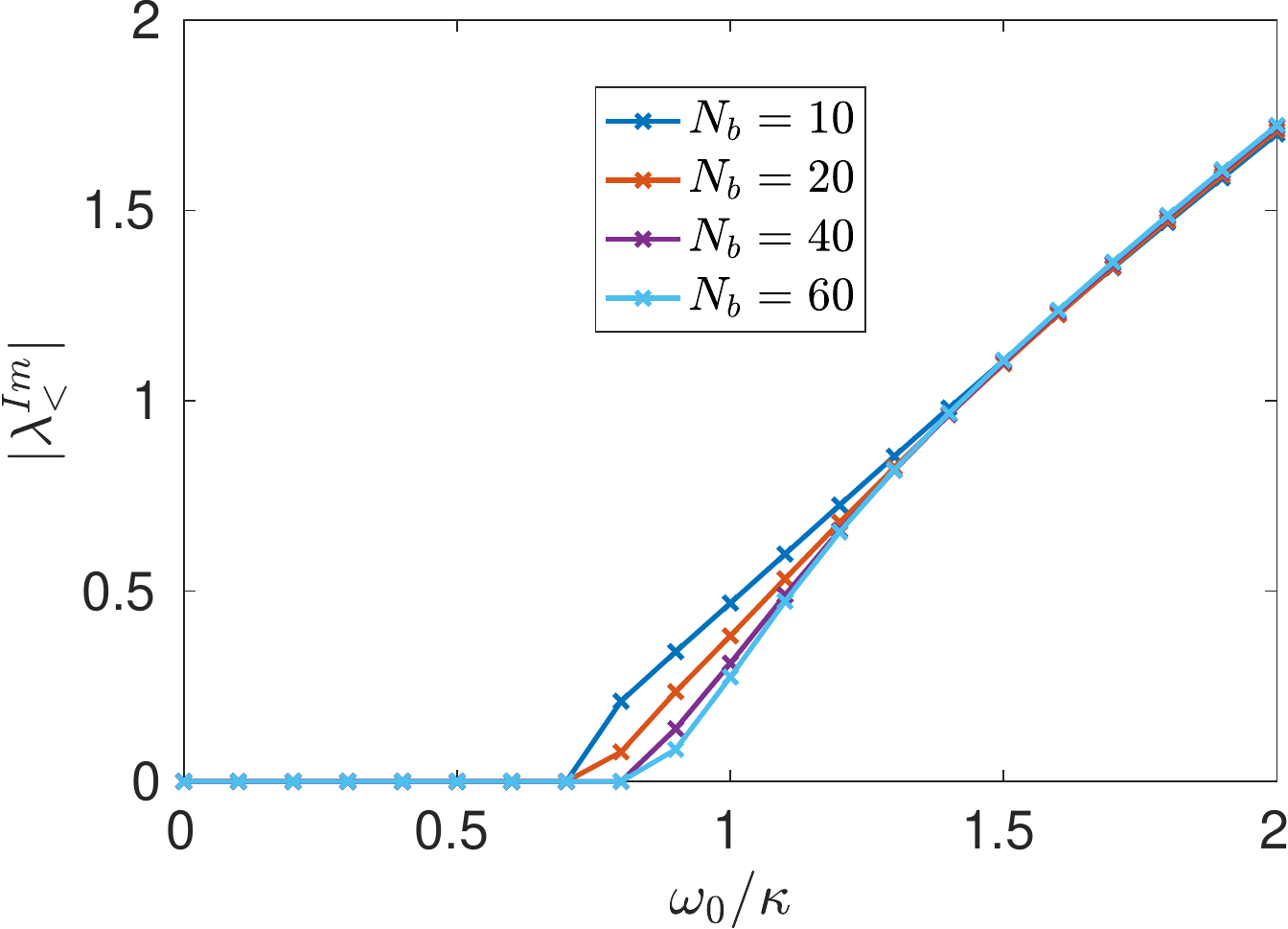}
 \caption{
 Lowest excitation 
 eigenvalue of the Liouvillian with nonzero imaginary value.}
\label{fig.fss.liouv.spec.imagpart}
 \end{figure}

\section{Semiclassical limit} 
Using the cyclic property of the trace, we can write the derivative of the expectation of any operator $\hat{O}$ in the form
 \begin{align}
  \frac{\ud}{\ud t}\mean{\hat{O}}&=\frac{\ud}{\ud t}\Tr(\hat{O}\hat{\rho}) \nonumber \\
  &= i\Tr([\hat H_b,\hat{O}]\hat{\rho}) 
  +\frac{\kappa}{2S}\Tr\left(([\hat{S}_+,\hat{O}]\hat{S}_-+\hat{S}_+[\hat{O},\hat{S}_-])\hat{\rho}\right)\,.
\end{align}
Taking for $\hat{O}$ the three spin components and using the commutation relations $[\hat{S}^\alpha,\hat{S}^\beta]=i\epsilon^{\alpha\beta\gamma}\hat{S}^\gamma$, we find
%
\begin{align} \label{eqmed:eqn}
  \frac{\ud}{\ud t}\mean{\hat{S}^x}=&-\frac{\omega_z}{S}\left(\mean{\hat{S}^y\hat{S}^z}+\mean{\hat{S}^z\hat{S}^y}\right)+\frac{\kappa}{2S}\left(\mean{\hat{S}^x\hat{S}^z}+\mean{\hat{S}^z\hat{S}^x}+\mean{\hat{S}^x}\right) \nonumber\\
  \frac{\ud}{\ud t}\mean{\hat{S}^y}=&-\omega_0\mean{\hat{S}^z}+\frac{(\omega_z - \omega_x)}{S}\left(\mean{\hat{S}^x\hat{S}^z}+\mean{\hat{S}^z\hat{S}^x}\right)+\frac{\kappa}{2S}\left(\mean{\hat{S}^y\hat{S}^z}+\mean{\hat{S}^z\hat{S}^y}-\mean{\hat{S}^y}\right)  \\
  \frac{\ud}{\ud t}\mean{\hat{S}^z}=&\omega_0\mean{\hat{S}^y}-\frac{\kappa}{S}\left(\mean{(\hat{S}^x)^2}+\mean{(\hat{S}^y)^2}+\mean{\hat{S}^z}\right)
 + \frac{\omega_x}{S} \left( \mean{\hat{S}^x\hat{S}^y} + \mean{\hat{S}^y\hat{S}^x} \right)  \,. \nonumber
\end{align}
%
On approaching the thermodynamic limit $S\to\infty$ it is convenient to define the reduced operators $\hat{m}^\alpha=\hat{S}^\alpha/S$. From the spin 
commutation relations we easily see that $[\hat{m}^\alpha,\hat{m}^\beta]=i\epsilon^{\alpha\beta\gamma}\hat{m}^\gamma/S$: in the limit $S\to\infty$ 
these operators commute, therefore  $\mean{\hat{m}^z\hat{m}^x}\simeq \mean{\hat{m}^z}\mean{\hat{m}^x}$. We can therefore write Eqs.~\eqref{eqmed:eqn} as
\begin{align} \label{eqmed1:eqn}
  \frac{\ud}{\ud t}m^x=&-2\omega_z\, m^ym^z+{\kappa}m^xm^z \nonumber\\
  \frac{\ud}{\ud t}m^y=&2(\omega_z-\omega_x)\, m^xm^z-\omega_0m^z+{\kappa}m^ym^z     \\
  \frac{\ud}{\ud t}m^z=&\omega_0m^y-{\kappa}\left((m^x)^2+(m^y)^2\right)
   + 2\omega_x m^x m^y  \,. \nonumber
\end{align}
up to corrections of order $1/S$. These equations conserve the norm $\mathcal{N}\equiv (m^x)^2+(m^y)^2+(m^z)^2$. 
When  $\omega_z=\omega_x=0$ there 
is another conserved quantity
\begin{equation} \label{conserved:eqn}
  \mathcal{M}\equiv\frac{m^x}{m^y-\omega_0/\kappa}\,.
\end{equation}
It is easy to see that this quantity is conserved by comparing the first two equations of Eq.~\eqref{eqmed1:eqn}, where  we find $\frac{\ud }{\ud t}\log m^x=\frac{\ud}{\ud t}\log \left(m^y-\frac{\omega_0}{\kappa}\right)$. Also for $\omega_z,\,\omega_x\neq0$ it is possible to show that there is another conserved quantity beyond the norm. Let us consider first, for simplicity, the case in which $\omega_x=0$. We can see that there is a conserved quantity which generalizes the one Eq.~\eqref{conserved:eqn} and has the form
\begin{align} \label{conserved1:eqn}
  \mathcal{R}_{\omega_z} &= 2\omega_z\log\left((\kappa m^y+2\omega_z m^x-\omega_0)^2+(\kappa m^x-2\omega_z m^y)^2\right)\nonumber\\
        &+2\kappa\atan\left(\frac{\kappa m^x-2\omega_z m^y}{\kappa m^y+2\omega_z m^x-\omega_0}\right)+2\kappa\pi n
%
\end{align}
(details on the derivation of $\mathcal{R}_{\omega_z}$ are given in the section below). As 
the expression of the conserved quantity involves the
  arctangent, one may notice that this conserved quantity has multiple branches.
  i.e. the quantity is defined only up to integer multiples of
  $2\kappa\pi$.  This means that on crossing the branch cut, $\kappa m^y +
  2\omega_z m^x - \omega_0=0$, the conserved quantity can switch branch (add
  or subtract a multiple of of $2\pi \kappa$).  As a result, the
  conservation of $\mathcal{R}_{\omega_z}$ is consistent with a converging spiral
  toward the fixed point. When $\omega_z=0$, $\mathcal{R}_{\omega_z}$ reduces to $\mathcal{R}_{0}=2\kappa\atan(\mathcal{M})$, consistently with our previous finding.
Therefore, whichever the parameters of the system, there exist two conserved quantities, $\mathcal{N}$ and $\mathcal{R}_{\omega_z}$.  This property motivates the fact that this system never shows a single periodic attractor but can show closed periodic orbits in some regimes  
(in other regimes the attractor can be a fixed point, as demonstrated in Ref.~\cite{hannukainen_2017_SI} 
for $\frac{\omega_0}{\kappa}<1$ and $\omega_z=0$). A case in which the system can only show closed periodic orbits is $\omega_z=0$ and $\frac{\omega_0}{\kappa}>1$, the weak dissipation regime we extensively consider in the main text. We can see this fact in the phase space portrait in the upper-left panel of Fig.~\ref{fig:single}. For the phase space portraits we consider the coordinates $\mathcal{Q}$ and $\mathcal{P}$ defined as
\begin{align}
 \mathcal{Q}&=m^z\nonumber\\
 m^x&=\sqrt{1-Q^2}\cos(2\mathcal{P})\\
 m^y&=\sqrt{1-Q^2}\sin(2\mathcal{P})\,.\nonumber
\end{align}
In the other panels we show the phase space portraits for different parameters. In the upper right panel  we consider $\omega_0/\kappa>1$ and $\omega_z$ small; we see that the dynamics is constrained over closed periodic orbits also in this case. Consistently with the existence of the conserved quantity Eq.~\eqref{conserved1:eqn}, the time-translation symmetry breaking is a phenomenon robust to this kind of perturbation of the Hamiltonian. 

On increasing $\omega_z$ there is a transition to a different regime where in half of the phase space there are closed orbits, and in the other half there is a single-point attractor (lower panel of Fig.~\ref{fig:single}). We can understand this fact from an analytical point of view, looking for the fixed points of Eq.~\eqref{eqmed1:eqn}. Imposing the time-derivatives equal to 0, {and $\mathcal{N}=1$, we find four fixed points: the trivial pair ($m^z=0, m^y=\kappa/\omega_0, m^x=\pm\sqrt{1-(\kappa/\omega_0)^2}$), and} two non-trivial ones at
\begin{align}
  m^x&=\frac{2\omega_z \omega_0}{\sqrt{\kappa^2+4\omega_z^2}}\nonumber\\
  m^y&=\frac{\kappa \omega_0}{\sqrt{\kappa^2+4\omega_z^2}}\\
  m^z&=\pm\sqrt{1-\frac{\omega_0^2}{\sqrt{\kappa^2+4\omega_z^2}}}\,.\nonumber
\end{align}
In order to find the non-trivial fixed points, we have to impose the argument of the square-root larger than 0, which gives the condition $\sqrt{\kappa^2+\omega_z^2}\geq \omega_0^2$. With the parameters used in Fig.~\ref{fig:single}, the transition point is at $\omega_z=\frac{\sqrt{3}}{4}$, which agrees with the numerical observations. As we can see in the lower panels of Fig.~\ref{fig:single}, one of the two fixed points is attractive (the one on the left) and the other is repelling  (the one on the right).

 In the case where $\omega_x \neq 0$ it is also possible to derive a conserved quantity for the dynamics similar to Eq.\eqref{conserved1:eqn} (see the details for its derivation in Section below). In Fig.\eqref{fig.phase.space.wx} we show the phase space portrait for the dynamics of the system, where we see that the dynamics is also constrained over closed periodic orbits, corroborating our expectations on the stability of the time crystal phase.
 
It is worth noticing that the peculiar dynamics studied in this article belongs to 
the class of reversible systems~\cite{Roberts_1992}, i.e., 
dynamical systems whose phase space variable ``$x$'' is invariant under the combination $t\rightarrow - t$, $x \rightarrow Gx$, with $G$ representing an involution transformation ($G \circ G$ = Identity).
From our equations of motion (Eq.~\eqref{eqmed1:eqn}), we promptly identify the involution
in our model as the transformation $m_x \rightarrow m_x$, 
$m_y \rightarrow m_y$ and $m_z \rightarrow -m_z$.

\section{ Derivation of conserved quantities $\mathcal{R_{\omega_{{\rm z}},\omega_{{\rm x}}}}$ } \label{below:sec}
 In order to construct a conserved quantity for $\omega_z$ and $\omega_x\neq 0$, with $\omega_z>\omega_x$, we rewrite the first two lines of Eq.~\eqref{eqmed1:eqn} as
 \begin{equation}\label{eq1:derivationR}
 \frac{1}{m^z}\frac{\ud}{\ud t}\left(\begin{array}{c}m^x\\m^y\end{array}\right)=
\hat A \left(\begin{array}{c}m^x\\m^y\end{array}\right)+
\left(\begin{array}{c}0\\-\omega_0\end{array}\right)\,.
\end{equation} 
with 
\begin{equation}
\hat A = \left(\begin{array}{cc}\kappa&-2\omega_z\\2(\omega_z - \omega_x)&\kappa\end{array}\right)
\end{equation}
Diagonalizing the $\hat A$ matrix we obtain its eigenvalues and eigenvectors, given respectively by,
\begin{equation}
\lambda_{\pm} = \kappa \pm 2i \sqrt{\omega_z(\omega_z-\omega_x)},\qquad 
\hat u_\pm = \frac{1}{\sqrt{c}} \left(\begin{array}{c} 1 \\ \mp i \sqrt{ 1-\omega_x/\omega_z} \end{array}\right)
\end{equation}
with $c = {2-\omega_x/\omega_z}$ the normalization constant.
We can now rewrite Eq.\eqref{eq1:derivationR} in this eigenbasis, obtaining
\begin{equation} \label{trasfo:eqn}
\frac{\ud}{\ud t} \hat \eta_\pm = \left( \lambda_\pm \hat \eta_\pm - \frac{\omega_0}{\sqrt{c}} \right) m^z
\end{equation}
where $\hat \eta_\pm = \hat u_\pm \cdot (m_x,m_y)^T$. In other words, we have that,
\begin{equation}
\frac{1}{\lambda_{\pm}} \frac{\ud}{\ud t} \log \left( \lambda_\pm \hat \eta_\pm - \frac{\omega_0}{\sqrt{c}} \right) = m^z
\end{equation}
From this last relation we directly find the conserved quantity,
\begin{equation}
\mathcal{R}_{\omega_z,\omega_x} \equiv 
-i\left[\lambda_- \log \left( \lambda_+ \hat \eta_+ - \frac{\omega_0}{\sqrt{c}} \right)
 - \lambda_+ \log \left( \lambda_- \hat \eta_- - \frac{\omega_0}{\sqrt{c}} \right)\right]
\end{equation}
which is clearly real being $\lambda_+=\lambda_-^*$. Considering, for example, the case with $\omega_x=0$, we can expand the terms in the conserved quantity as,
\begin{align} \label{conserved2:eqn}
  \mathcal{R}_{\omega_z} = (-i\kappa+2\omega_z)\log\left(im^x+m^y-\frac{\omega_0}{(\kappa-2i\omega_z)}\right)
   -(-i\kappa-2\omega_z)\log\left(-im^x+m^y-\frac{\omega_0}{(\kappa+2i\omega_z)}\right)\,.
\end{align}
in which after some straightforward formal manipulations we get Eq.~\eqref{conserved1:eqn}, up to an immaterial constant which we do neglect. Due to the branch cut of the logarithms, the  quantity in Eq.~\eqref{conserved1:eqn} is defined only up to integer multiples of  $2\kappa\pi$.  This offset is determined by the number of times the branch cut is crossed. We show an instance of this fact in Fig.~\ref{branch-cut:fig}, where we consider the case of trajectories spiraling towards the attracting fixed point. The solid blue line marks the branch cut, the red lines mark different trajectories with different values of $\mathcal{R}_{\omega_z}$: each time a trajectory crosses the branch cut, the value of $\mathcal{R}_{\omega_z}$ increases by $2\pi \kappa$.

\begin{figure*}
\begin{center}
\begin{tabular}{cc}
\hspace{-1cm}\includegraphics[width=8cm]{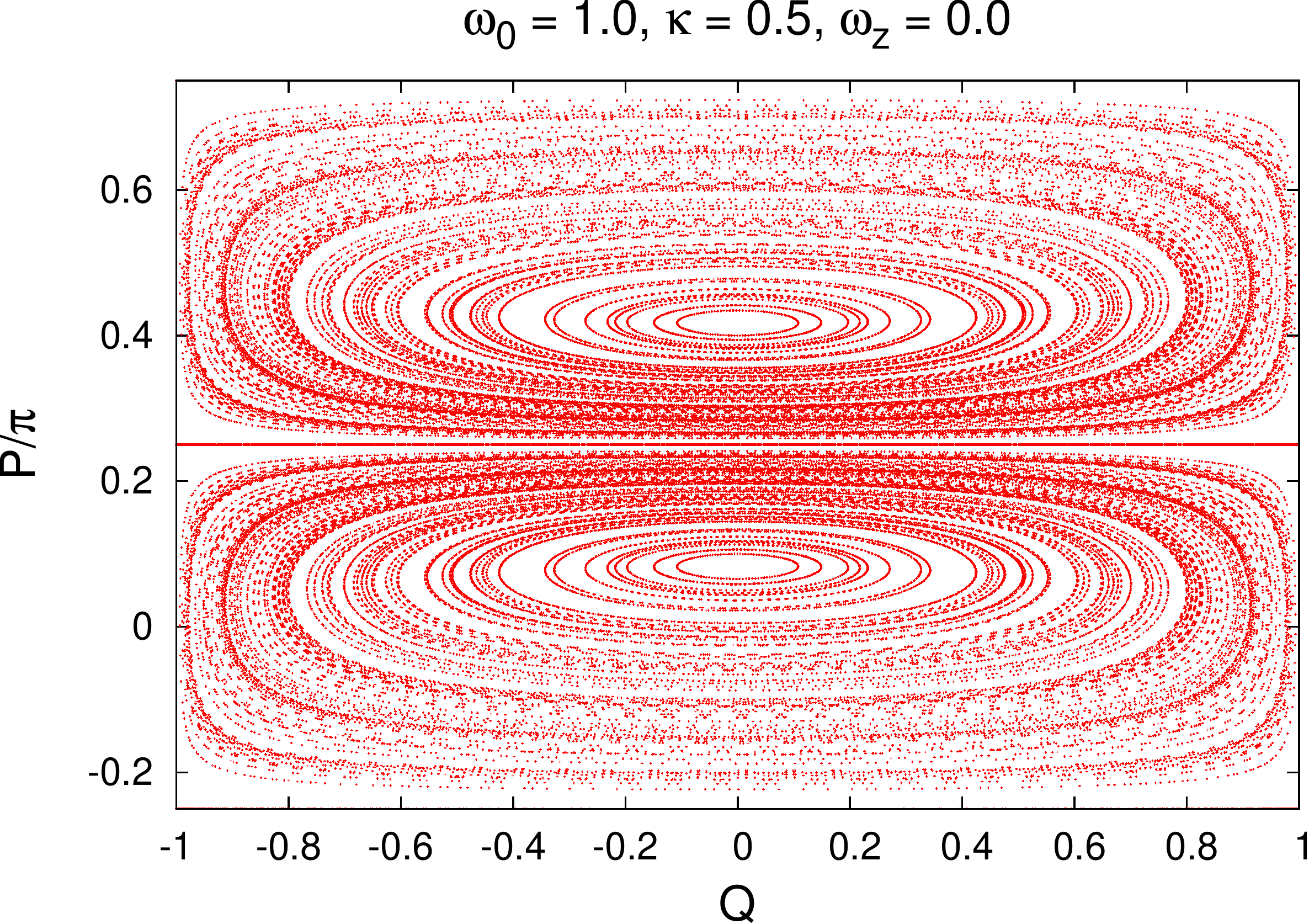}&
\hspace{0.3cm}\includegraphics[width=8cm]{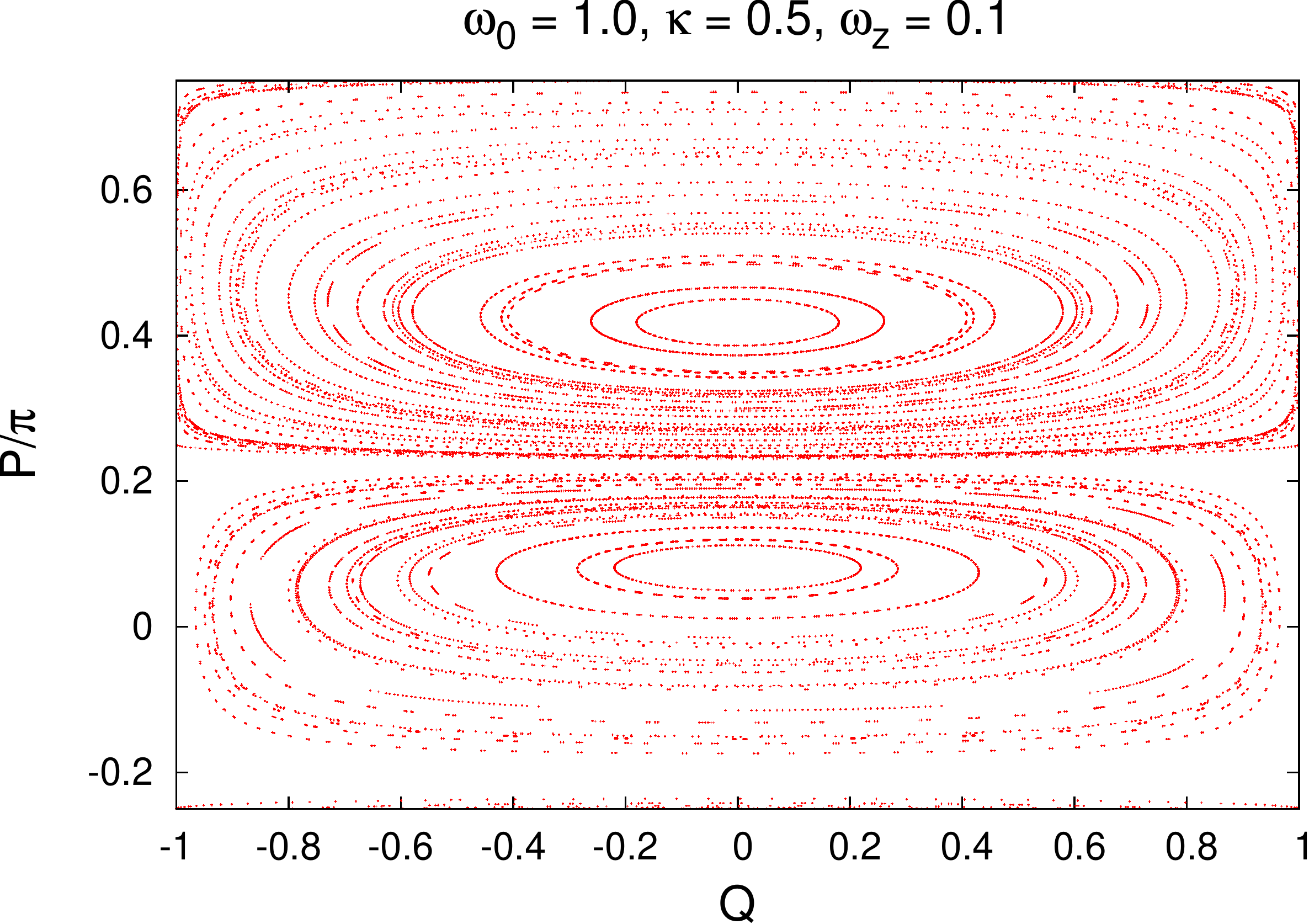}\\
\\
\hspace{-1cm}\includegraphics[width=8cm]{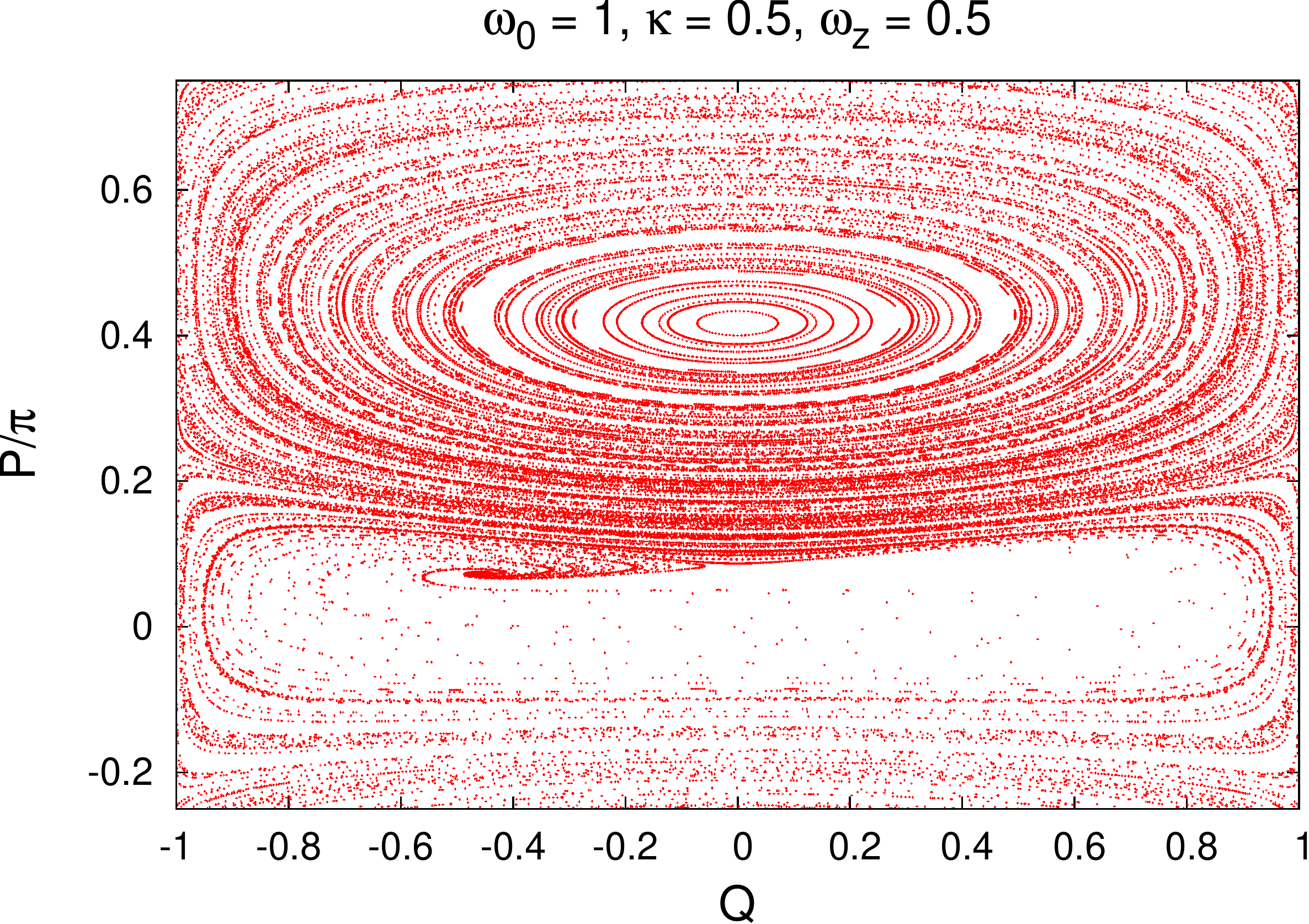}&
\hspace{0.3cm}\includegraphics[width=8cm]{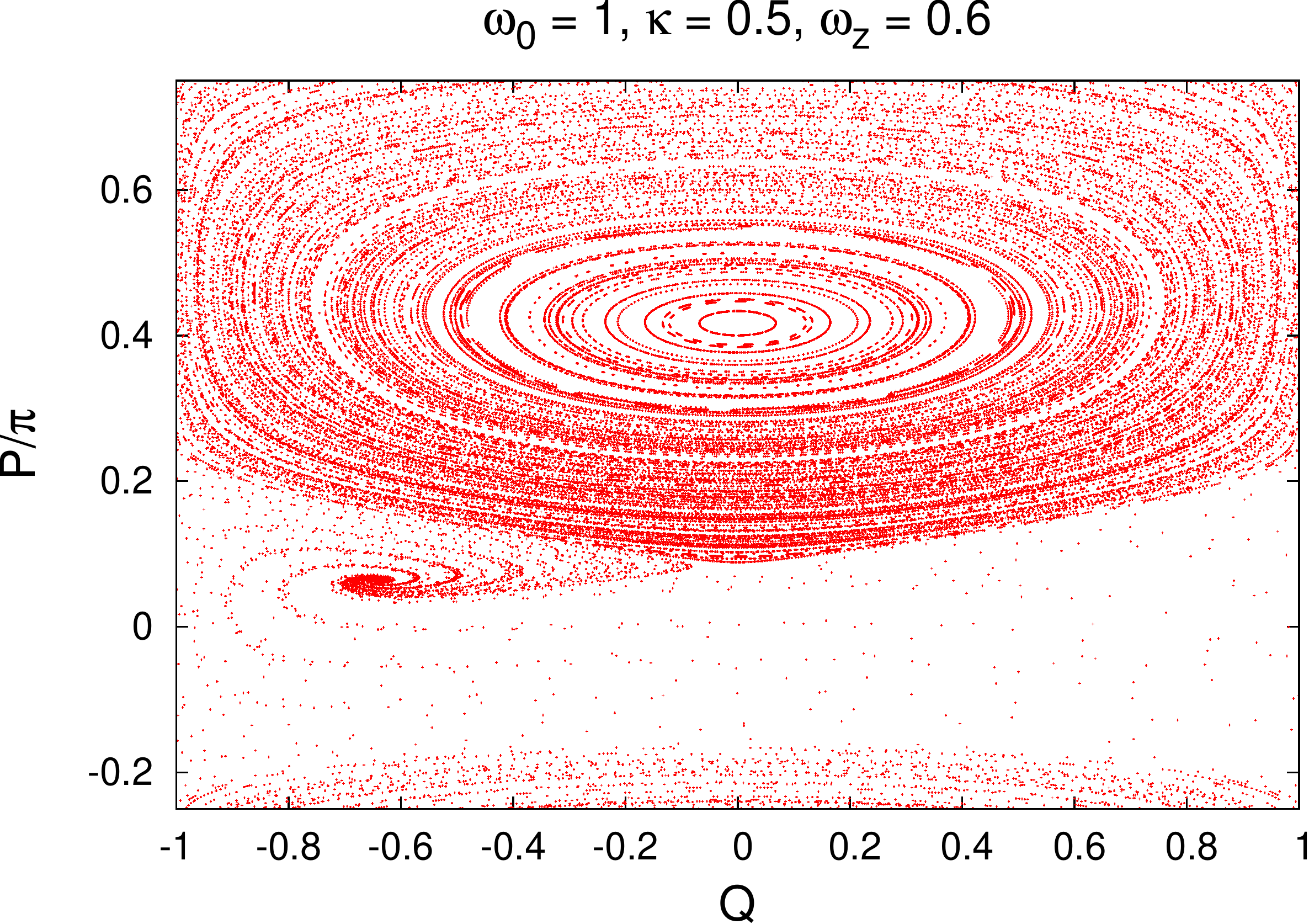}
\end{tabular}
\centering
\caption{ Phase space portraits for the dynamics in Eqs.~\eqref{eqmed1:eqn}, with $\omega_x=0$, for different values of the system parameters. We see that for small values of $\omega_z$ the dynamics is constrained to closed periodic orbits, while for larger values there is a portion of phase space where initial conditions are attracted towards a stable fixed point. 
}
\label{fig:single}
\end{center}
\end{figure*} 
  \begin{figure}
 \centering
  \includegraphics[width=0.45\textwidth]{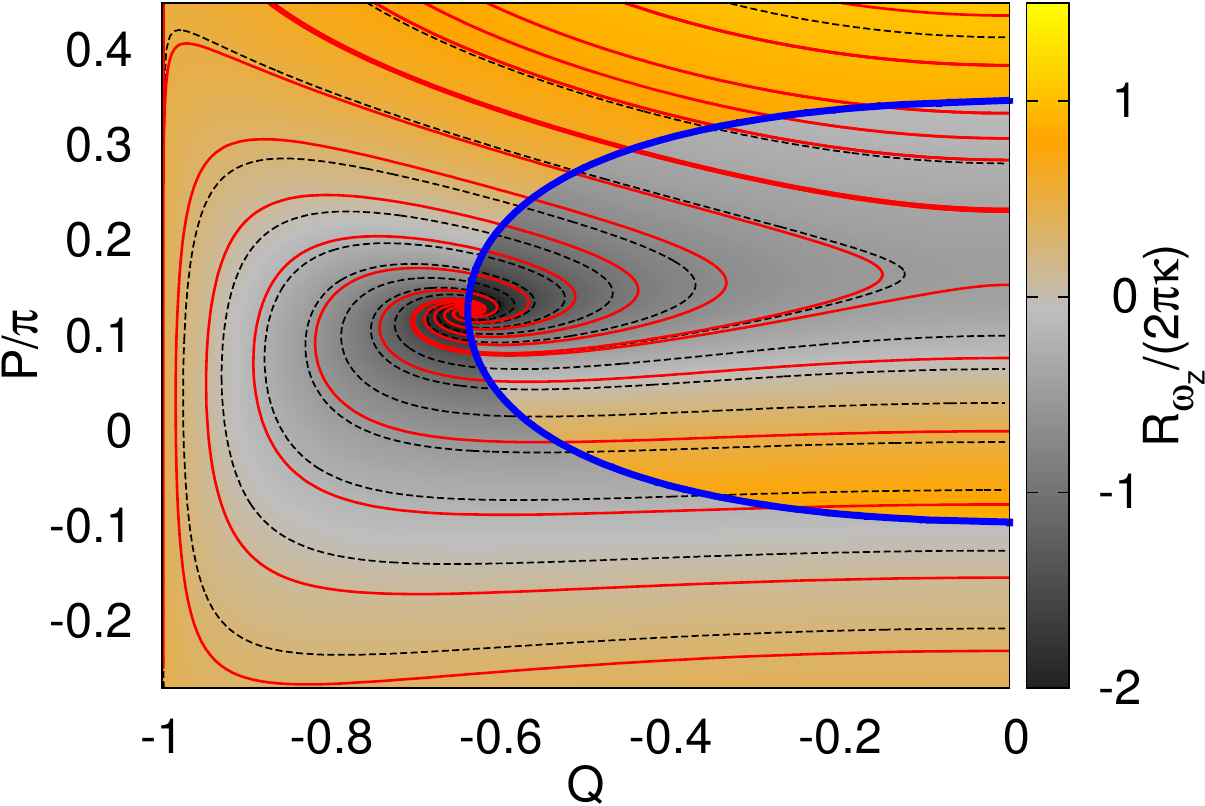}
 \caption{Trajectories (red solid) spiraling towards the fixed point  and the branch cut of the difference of the logarithms in Eq.~\eqref{conserved2:eqn} (blue solid line).  The color-scale and gray-dashed contours correspond to the conserved quantity $\mathcal{R}/2\pi \kappa$ (Numerical parameters: $\omega_0/\kappa=2.0,\,\omega_z/\kappa=1.2$).}
\label{branch-cut:fig}
 \end{figure}
 \begin{figure}
 \centering
 \includegraphics[width=0.48\textwidth]{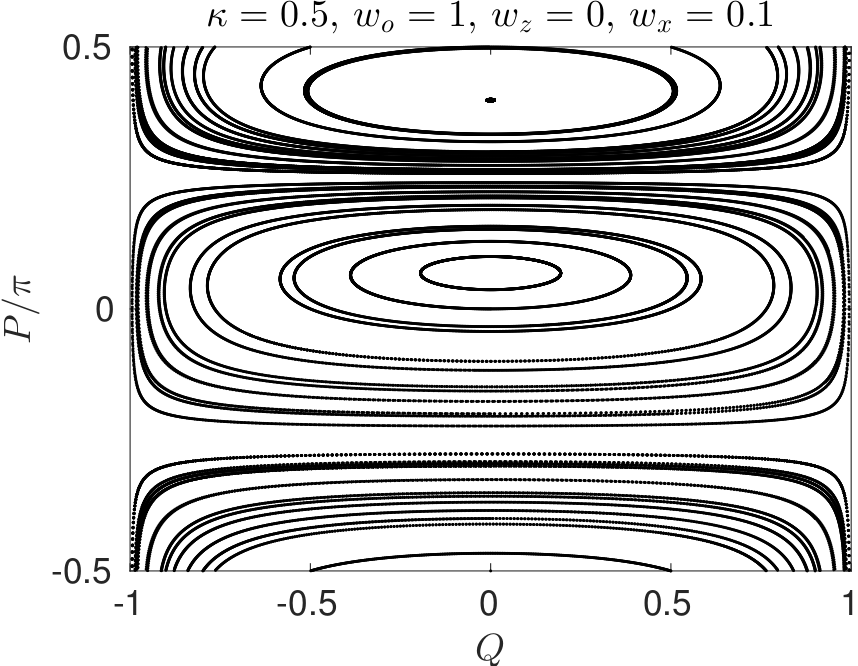}
 \includegraphics[width=0.48\textwidth]{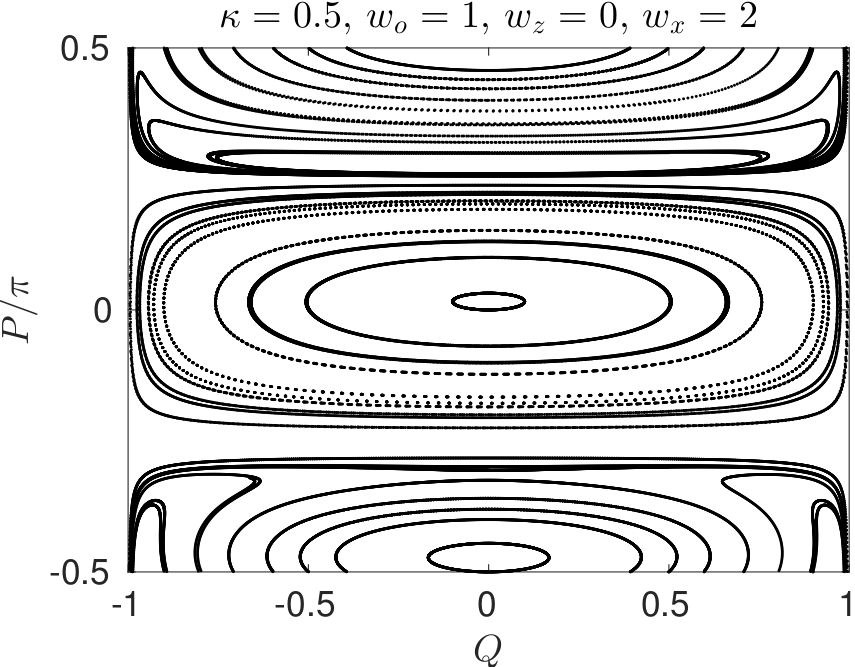}\\
 \includegraphics[width=0.48\textwidth]{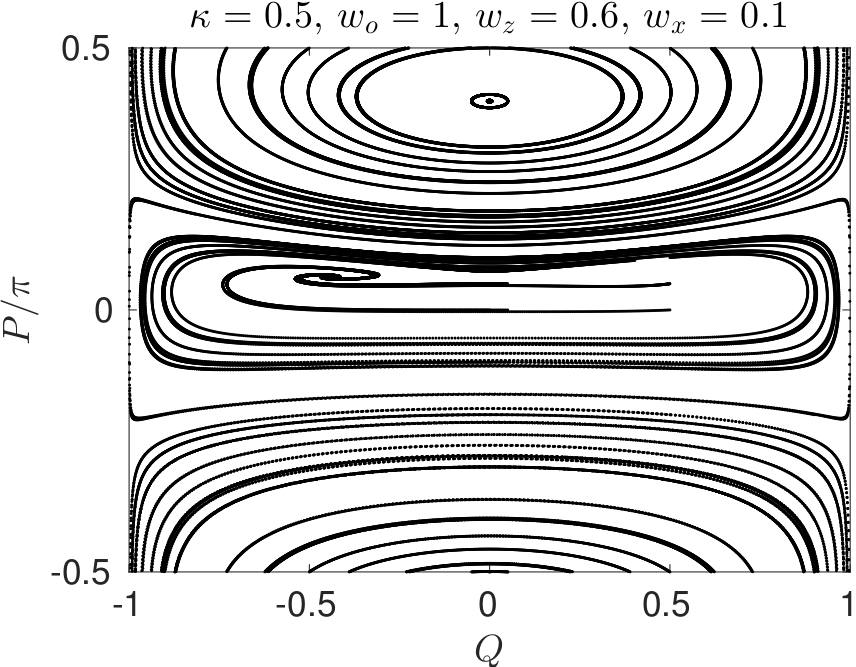}
 \includegraphics[width=0.48\textwidth]{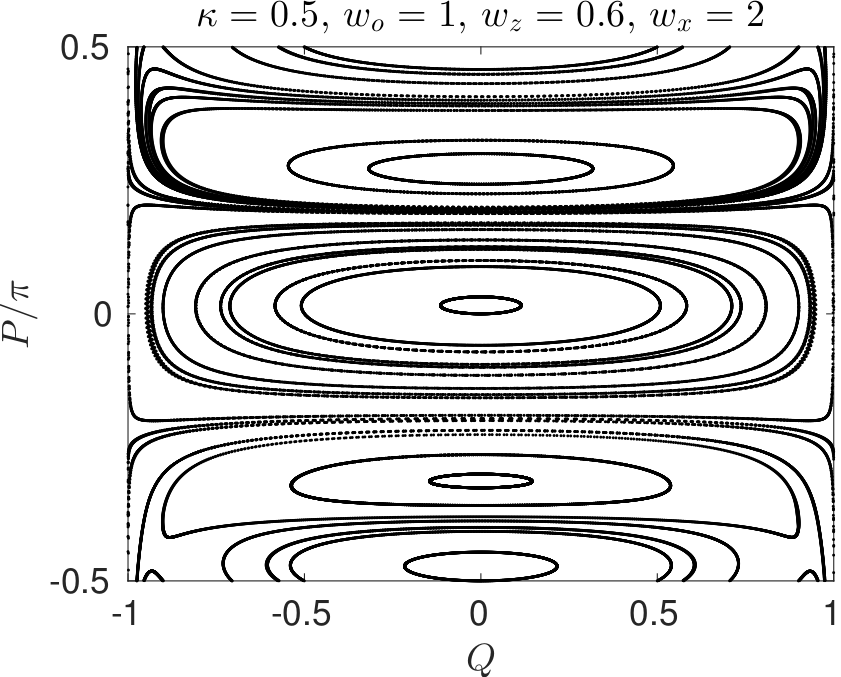}
 \caption{  Phase space portraits for the dynamics in Eqs.~\eqref{eqmed1:eqn}, with $\omega_x \neq 0$,  for different values of the system parameters. We see that even for large values of $\omega_x$ the dynamics is still constrained to closed periodic orbits. In fact, such term improves the stability of the time crystal: the attractive fixed point in the lower left-panel stabilizes to periodic orbits (lower right-panel) at larger values of $\omega_x$.
}
 \label{fig.phase.space.wx}
 \end{figure}

\end{document}